\documentclass[usenatbib]{mn2e}

\usepackage{epsf}

\newcommand\be{{\bmath e}}
\newcommand\bff{{\bmath f}}
\newcommand\bn{{\bmath n}}
\newcommand\br{{\bmath r}}
\newcommand\bu{{\bmath u}}
\newcommand\bv{{\bmath v}}
\newcommand\bS{{\bmath S}}
\newcommand\bOmega{{\bmath\Omega}}
\newcommand\bnabla{{\bmath\nabla}}

\newcommand\rmc{\mathrm{c}}
\newcommand\rmd{\mathrm{d}}
\newcommand\rme{\mathrm{e}}

\newcommand\rmg{\mathrm{g}}
\newcommand\rmh{\mathrm{h}}
\newcommand\rmi{\mathrm{i}}
\newcommand\rmnw{\mathrm{nw}}
\newcommand\rms{\mathrm{s}}
\newcommand\rmt{\mathrm{t}}

\newcommand\rmw{\mathrm{w}}

\newcommand\f{\frac}
\newcommand\p{\partial}
\newcommand\cst{\mathrm{constant}}

\title[Tides in rotating barotropic fluid bodies] {Tides in rotating
  barotropic fluid bodies: the contribution of inertial waves and the
  role of internal structure}

\author[Gordon I. Ogilvie]{Gordon I. Ogilvie\\
Department of Applied Mathematics and Theoretical Physics,
University of Cambridge, Centre for Mathematical Sciences,\\
Wilberforce Road, Cambridge CB3 0WA
}

\begin{document}

\maketitle

\label{firstpage}
 
\begin{abstract}
  We discuss the linear response to low-frequency tidal forcing of
  fluid bodies that are slowly and uniformly rotating, are neutrally
  stratified and may contain a solid or fluid core.  This problem may
  be regarded as a simplified model of astrophysical tides in
  convective regions of stars and giant planets.  The response can be
  separated into non-wavelike and wavelike parts, where the former is
  related instantaneously to the tidal potential and the latter may
  involve resonances or other singularities.  The imaginary part of
  the potential Love number of the body, which is directly related to
  the rates of energy and angular momentum exchange in the tidal
  interaction and to the rate of dissipation of energy, may have a
  complicated dependence on the tidal frequency.  However, a certain
  frequency-average of this quantity is independent of the dissipative
  properties of the fluid and can be determined by means of an impulse
  calculation.  The result is a strongly increasing function of the
  size of the core when the tidal potential is a sectoral harmonic,
  especially when the body is not strongly centrally condensed.
  However, the same is not true for tesseral harmonics, which receive
  a richer response and may therefore be important in determining
  tidal evolution even though they are usually subdominant in the
  expansion of the tidal potential.  We also discuss analytically the
  low-frequency response of a slowly rotating homogeneous fluid body
  to tidal potentials proportional to spherical harmonics of degrees
  less than five.  Tesseral harmonics of degrees greater than two,
  such as are present in the case of a spin-orbit misalignment, can
  resonate with inertial modes of the full sphere, leading to an
  enhanced tidal interaction.
 \end{abstract}

\begin{keywords}
  hydrodynamics -- waves -- planets and satellites: general -- planets and satellites: interiors -- planet--star interactions -- binaries: close
\end{keywords}

\section{Introduction}

The tidal interaction of two bodies orbiting about their centre of
mass is one of the fundamental problems of theoretical astrophysics.
It is relevant to a wide variety of systems, including close binary
stars, the satellites of solar-system bodies, and stars orbiting the
black holes in galactic centres.  Interest in this subject has grown
through the discovery in recent years of numerous planets in orbits of
a few days or less around other stars.\footnote{See
  \texttt{exoplanet.eu} or \texttt{exoplanets.org}.}  It is likely
that tidal interactions have affected the orbital and spin properties
of many of these planets as well as providing a source of heat.
Equally importantly, they may also have led to the destruction of many
planets that are not observed.

Tidal theory involves a combination of celestial mechanics and
continuum mechanics.  The celestial mechanics is non-trivial, and has
been widely investigated using highly simplistic descriptions or
parametrizations of the fluid or solid behaviour
\citep[e.g.][]{1880RSPT..171..713D,1961MNRAS.122..339J,1963MNRAS.126..257G,1966Icar....5..375G,1973Ap&SS..23..459A,1980M&P....23..185M,1981A&A....99..126H,1998ApJ...499..853E,2002ApJ...573..829M,2009MNRAS.395.2268B}.
In fact, the fluid dynamics of tidally forced bodies is a much richer
and more difficult problem, even in a linear theory where the tide is
treated as an infinitesimal disturbance.  In the case of stars, the
work of \citet[][and references therein]{1977A&A....57..383Z}
identified two potentially important processes: the interaction of
tidal bulges with turbulent convection and the excitation and damping
of internal gravity waves in radiative zones \citep[see
also][]{1983MNRAS.203..581S,1989ApJ...342.1079G,1998ApJ...502..788T,1998ApJ...507..938G,2010MNRAS.404.1849B}.
Rotation was neglected in this work in order to simplify the
calculations.  Its effects have been considered by
\citet{1999A&A...341..842W} and \citet{2002A&A...386..211S} within the
`traditional approximation', which is usually applicable to the
radiative zones of slowly rotating stars.

More recent work has examined the effects of the full Coriolis force
on tides in convective zones
(\citealt{2004ApJ...610..477O,2005ApJ...635..674W,2005ApJ...635..688W,2005MNRAS.364L..66P,2005JFM...543...19O,2007ApJ...661.1180O,2009ApJ...696.2054G,2009MNRAS.396..794O,2010JFM...643..363R};
see also
\citealt{1995MNRAS.277..471S,1997MNRAS.291..633S,1997MNRAS.291..651P}).
All of these papers show that inertial waves, for which the Coriolis
force provides the restoring force, can be excited by tidal forcing
and may provide an important route for tidal dissipation in rotating
bodies.  However, even the linear theory of inertial waves is a very
intricate problem and these authors come to different conclusions.  In
some cases the response is dominated by wave attractors, critical
latitudes or other singularities, while in others global normal modes
are excited.  Different predictions are made for the
frequency-dependence of the tidal dissipation rate.  For example,
while \citet{2004ApJ...610..477O}, \citet{2009MNRAS.396..794O} and
\citet{2010JFM...643..363R} allow inertial waves to undergo multiple
reflections within a spherical shell before being damped by viscosity,
and obtain a highly frequency-dependent dissipation rate,
\citet{2009ApJ...696.2054G} do not allow any reflections of waves
after their generation from the boundary of the planetary (or stellar)
core, but instead consider nonlinear damping and obtain a smooth
dissipation rate that increases with the fifth power of the core size.
Nevertheless, \citet{2009MNRAS.396..794O} found that the
frequency-averaged dissipation rate with multiple reflections shows
the same dependence on core size.

One purpose of this paper is to show that, in spite of this
complexity, some broader aspects of the tidal response of rotating
bodies are robust and insensitive to the details that distinguish
between these calculations.  Even if the linear theory of inertial
waves in a perfect spherical shell cannot be applied reliably to the
turbulent convective zone of a star or planet, the results presented
here on frequency-integrated responses should still be valid.

A second purpose of this paper is to collect some analytical results
on the tidal response of a homogeneous fluid body.  When potential
components other than the usual $l=m=2$ spherical harmonic are
considered, a richer response is possible and numerous resonances can
occur with large-scale inertial modes, especially in systems with a
spin-orbit misalignment.  Similar behaviour can be expected in more
realistic models, and the astrophysical consequences require further
investigation.

\section{Spheroidal and toroidal vector fields}

We begin with a definition that is important for the analysis in this
paper.  Let $(r,\theta,\phi)$ be spherical polar coordinates, and
$\be_r=\br/r$ the radial unit vector.  A general differentiable vector
field $\bv$ can be represented in the form
\begin{equation}
  \bv=\be_r\,R-\be_r\times(\be_r\times\bnabla S)-\be_r\times\bnabla T,
\end{equation}
where $R$, $S$ and $T$ are scalar functions of $(r,\theta,\phi)$.  Thus $\bv=\bv_\rms+\bv_\rmt$, where the spheroidal and toroidal parts of $\bv$ are
\begin{equation}
  \bv_\rms=\be_r\,R-\be_r\times(\be_r\times\bnabla S),
\end{equation}
\begin{equation}
  \bv_\rmt=-\be_r\times\bnabla T.
\end{equation}
This nomenclature is used, for example, in seismology
\citep[e.g.][]{Lapwood1981} and stellar oscillations
\citep[e.g.][]{Smeyers2010}, and is related to the theory of vector
spherical harmonics \citep[e.g.][]{Morse1953}.  We have
\begin{equation}
  \be_r\cdot\bv=R,
\label{r}
\end{equation}
\begin{equation}
  \be_r\cdot\bnabla\times(-\be_r\times\bv)=-\nabla_\rmh^2S,
\end{equation}
\begin{equation}
  \be_r\cdot\bnabla\times\bv=-\nabla_\rmh^2T,
\label{t}
\end{equation}
where
\begin{equation}
  \nabla_\rmh^2=\f{1}{r^2\sin\theta}\f{\p}{\p\theta}\left(\sin\theta\f{\p}{\p\theta}\right)+\f{1}{r^2\sin^2\theta}\f{\p^2}{\p\phi^2}
\end{equation}
is the horizontal part of the Laplacian operator.  On each sphere
$r=\cst>0$, the operator $\nabla_\rmh^2$ has a complete set of
eigenfunctions (spherical harmonics) and can be inverted uniquely
except for the addition of an arbitrary constant.  Therefore, given a
differentiable vector field $\bv$, equations (\ref{r})--(\ref{t})
determine $R$, $S$ and $T$ uniquely except that arbitrary functions of
$r$ only can be added to $S$ and $T$.

In \citet{2004ApJ...610..477O} and \citet{2009MNRAS.396..794O} the
velocity field is separated into spheroidal and toroidal parts and
expanded in spherical harmonics.  The coefficients $a_n$ and $b_n$
refer to the spheroidal part ($R$ and $S$), while the coefficients
$c_n$ refer to the toroidal part ($T$).  In
\citet{2009MNRAS.396..794O}, for example, we have
\begin{equation}
  R=\sum_na_n(r)Y_n^m(\theta,\phi),
\end{equation}
\begin{equation}
  S=r^2\sum_nb_n(r)Y_n^m(\theta,\phi),
\end{equation}
\begin{equation}
  T=r^2\sum_nc_n(r)Y_n^m(\theta,\phi),
\end{equation}
where the sums are over integers $n\ge m\ge0$,
\begin{equation}
  Y_n^m(\theta,\phi)=\left[\f{(2n+1)(n-m)!}{4\pi(n+m)!}\right]^{1/2}P_n^m(\cos\theta)\,\rme^{\rmi m\phi}
\end{equation}
is a spherical harmonic with the normalization
\begin{equation}
  \int_0^{2\pi}\int_0^\pi|Y_n^m(\theta,\phi)|^2\,\sin\theta\,\rmd\theta\,\rmd\phi=1,
\end{equation}
and the real part of these expressions is to be taken after they are
multiplied by $\rme^{-\rmi\omega t}$.

\section{Tidal response of a homogeneous rotating body}

Perhaps the simplest tidal problem is the linear response of a
homogeneous fluid sphere to a tidal potential that is harmonic in time
and proportional to a solid spherical harmonic \citep[see][although
his analysis is mainly for an elastic solid rather than a
fluid]{Thomson1863}.  To include the effects of rotation, we should
properly consider an ellipsoidal figure of equilibrium
\citep{Chandrasekhar1969}; tides in a Maclaurin spheroid were analysed
by \citet{Bryan1889}, although the full ramifications of his work have
not yet been explored.  In the limit of slow rotation, however, it is
adequate to neglect centrifugal effects and consider a spherical fluid
body subject to the Coriolis force.

The tidal gravitational potential experienced by a body can be written
as an interior multipole expansion in solid spherical harmonics of
degrees $l\ge2$; this series converges more rapidly when the
tide-raising body is more distant.  For each value of $l$, the
response of a rotating body also depends on the order $m$ of the
spherical harmonic, which satisfies $-l\le m\le l$.  (The
response for negative values of $m$ can be deduced from that for
positive $m$ by complex conjugation if the frequency is
also changed in sign.)

We collect in Appendix~\ref{a:homogeneous} some analytical results on
the linear response of a homogeneous sphere to forcing by solid
spherical harmonics with degrees $l=\{2,3,4\}$, i.e.\ interior
quadrupolar, octupolar and hexadecapolar potentials.  (No tidal force
is generated by $l=0$ or $l=1$.)  This analysis makes use of the
decomposition of the tidal response into non-wavelike and wavelike
parts, developed in Section~\ref{s:asymptotics} below.

In general, the tidal response involves resonances with inertial modes
of the sphere when the tidal frequency matches the eigenfrequency of
an appropriate inertial mode.  The eigenfrequencies of inertial modes
in the frame that rotates with the angular velocity $\Omega$ of the
body are dense in the interval $(-2\Omega,2\Omega)$
\citep{1968Greenspan}.  However, the selection rules determined by the
spatial structure of the inertial modes and the forcing functions
imply that any spherical harmonic potential resonates with only a
finite number of inertial modes.  Indeed, as found in
Appendix~\ref{a:homogeneous}, the sectoral harmonics ($|m|=l$) do not
resonate with any inertial modes, while the tesseral harmonics
($|m|<l$) do have such resonances.

However, the resonances of the quadrupolar tesseral harmonics $Y_2^0$
and $Y_2^1$ are only formal resonances and should not lead to enhanced
dissipation in linear theory.  The formally resonant modes have zero
frequency in an inertial frame of reference; they are present because
of the existence of a continuous family of equilibrium solutions for
uniformly rotating bodies that have angular velocity vectors that
differ in magnitude and/or direction.  In the case $m=0$ the
zero-frequency mode is the `spin-up' mode, which consists simply of an
infinitesimal uniform change in the angular velocity of the fluid.  In
the case $m=1$ it is the `spin-over' mode of frequency
$\omega=-\Omega$ (i.e.\ zero frequency in the inertial frame), which
consists of an infinitesimal tilt of the rotation axis of the fluid.
Each mode involves a change in the exactly conserved angular momentum
of the body and therefore has zero frequency.  The formal resonances
of $Y_2^0$ and $Y_2^1$ with these modes could not produce any tidal
dissipation because the modes concerned involve uniform rotation and
the resonance occurs at zero forcing frequency (in the inertial
frame).  It may be surprising that the spin-up mode, which involves a
change in the angular velocity of the fluid, can be excited by an
axisymmetric tidal potential proportional to $Y_2^0$, which cannot
change the axial component of angular momentum of the body.  In fact,
even in the absence of a torque, the spin-up mode is excited because
the axisymmetric tidal deformation modulates the moment of inertia of
the body.

The most important harmonic present in the tidal potential is usually
considered to be $Y_2^2$, and the fact that it cannot resonantly
excite inertial normal modes in a homogeneous full sphere has been
noted previously \citep[e.g.][]{2009ApJ...696.2054G}.  However, given
that tesseral harmonics have a richer response in a homogeneous body,
and therefore presumably also in an inhomogeneous body, other
harmonics are worthy of consideration, even if they are subdominant in
the expansion of the tidal potential.

To illustrate this, we consider here the case of two bodies in a
circular orbit of angular velocity (or mean motion) $n$.  We consider
the tide raised by body~2 in body~1, which rotates uniformly, and
assume that the orbit is inclined with respect to the equatorial plane
of body~1.  [Related analyses are given by \citet{2009MNRAS.395.2268B}
and \citet{2012MNRAS.423..486L}.]  We consider four spherical polar
coordinate systems, all centred on body~1:
\begin{enumerate}
\item[$S_1$:] aligned with the orbit, rotating with the orbit
\item[$S_2$:] aligned with the orbit, non-rotating
\item[$S_3$:] aligned with the spin of body~1, non-rotating
\item[$S_4$:] aligned with the spin of body~1, rotating with body~1
\end{enumerate}
In $S_1$ the tide-raising body (as represented by its centre of mass)
is stationary and lies in the plane $\theta=\pi/2$.  The tidal
potential is independent of time and can be expanded in solid
spherical harmonics.  Let us focus on terms of a fixed degree
$l\ge2$.  Then harmonics of orders
$m=\{l,\,l-2,\,l-4,\,\dots,\,-l\}$ are present, because
symmetry about the plane $\theta=\pi/2$ requires that $l-m$ be
even.  In $S_2$ the tide-raising body executes a circular orbit in the
plane $\theta=\pi/2$.  The tidal potential involves the same spherical
harmonics but their angular frequencies are now $mn$.  The rotation
from $S_2$ to $S_3$ mixes all permissible values of $m$ for a given
$l$ (as described by the Wigner $D$-matrix).  In $S_3$, therefore,
all values of $m$ with $-l\le m\le l$ are present and the
frequencies are $\{l,\,l-2,\,l-4,\,\dots,\,-l\}n$ for each
$m$.  Finally, in $S_4$, all values of $m$ are present and the
frequencies are $\{l,\,l-2,\,l-4,\,\dots,\,-l\}n-m\Omega$,
where $\Omega$ is the spin angular velocity of body~1.

In Appendix~\ref{a:homogeneous} we show that, in $S_4$, forcing by
$Y_3^m$ resonates with an inertial mode of the sphere at frequencies
\begin{equation}
  \omega=\f{2}{3}\left[-m\pm\left(\f{9-m^2}{5}\right)^{1/2}\right]\Omega
\end{equation}
for $-2\le m\le2$.  Resonance with the octupolar tide therefore occurs
when
\begin{eqnarray}
  \lefteqn{\f{n}{\Omega}=\pm\{0,\,0.1700,\,0.2981,\,0.3922,\,0.4444,\,0.5099,}&\\\nonumber
  &&\qquad0.8944,\,1.1766,\,1.3333\}.
\end{eqnarray}
Similarly, forcing by $Y_4^m$ resonates with an inertial mode at
frequencies that satisfy the cubic equation
\begin{equation}
  42\left(\f{\omega}{\Omega}\right)^3+63m\left(\f{\omega}{\Omega}\right)^2-36(2-m^2)\f{\omega}{\Omega}-4m(11-2m^2)=0
\label{cubic}
\end{equation}
for $-3\le m\le3$.  Resonance with the hexadecapolar tide therefore
occurs when
\begin{eqnarray}
  \lefteqn{\f{n}{\Omega}=\pm\{0,\,0.0970,\,0.1770,\,0.1920,\,0.1940,\,0.25,}&\nonumber\\
  &&\qquad0.3273,\,0.3540,\,0.3840,\,0.4550,\,0.5,\nonumber\\
  &&\qquad0.5580,\,0.625,\,0.6547,\,0.9100,\,1.1160,\,1.25\}.
\end{eqnarray}
It is likely that these resonances persist, albeit with modifications,
in more realistic models of stars and giant planets.  The resonances
at smaller values of $n/\Omega$ may be expected to play a role in the
tidal synchronization of close binary stars and possibly of hot
Jupiters, although in the latter case tidal potential components
beyond the quadrupolar ones are intrinsically very weak because of the
large ratio of orbital semimajor axis to planetary radius.  The
resonances at larger values of $n/\Omega$, close to unity, could be
important for the more rapidly rotating hosts of hot Jupiters, such as
F~stars, where spin-orbit misalignments are common
\citep{2012arXiv1206.6105A}, although the convective zones of these
stars are of limited extent.

\section{Tidal forcing of a rotating barotropic fluid}

\subsection{Equilibrium}
\label{s:equilibrium}

A barotropic fluid is one in which the pressure $p$ is uniquely
related to the density $\rho$.  This situation arises, in particular,
when the specific entropy of the fluid does not vary in space or time.
The differential relations between the pressure, specific volume
$v=1/\rho$, specific internal energy $e$ and specific enthalpy
$h=e+pv$ are then\footnote{If the fluid is barotropic for some other
  reason, then $e$ and $h$ are not the thermodynamic internal energy
  and enthalpy, but play equivalent roles in the fluid dynamics.  Some
  heat exchange with the surroundings is then implied.} $\rmd
e=-p\,\rmd v$ and $\rmd h=v\,\rmd p$.  The sound speed is
$v_\mathrm{s}=\sqrt{\rmd p/\rmd\rho}$, so that $\rmd
h=v_\rms^2\,\rmd\rho/\rho$.

In a frame of reference that rotates with uniform angular velocity $\bOmega$, an ideal barotropic fluid satisfies the equation of motion,
\begin{equation}
  \f{\p\bu}{\p t}+\bu\cdot\bnabla\bu+2\bOmega\times\bu=-\bnabla(h+\Phi_\rmg+\Phi_\rmc),
\end{equation}
and the equation of mass conservation,
\begin{equation}
  \f{\p\rho}{\p t}+\bnabla\cdot(\rho\bu)=0,
\end{equation}
where $\bu$ is the velocity, $\Phi_\rmg$ is the gravitational potential, which satisfies Poisson's equation,
\begin{equation}
  \nabla^2\Phi_\rmg=4\pi G\rho,
\end{equation}
and
\begin{equation}
  \Phi_\rmc=-\f{1}{2}|\bOmega\times\br|^2
\end{equation}
is the centrifugal potential.

Steady axisymmetric solutions can be found in which the fluid is
uniformly rotating and therefore has $\bu=\mathbf{0}$ in the
appropriate frame of reference.  Such solutions satisfy the
equilibrium condition
\begin{equation}
  h+\Phi_\rmg+\Phi_\rmc=\cst.
\end{equation}

We assume that the barotropic relation is such that $p$ is an
increasing function of $\rho$ for all $\rho>0$, and that
$p\propto\rho^{1+1/n}$ in the limit $\rho\to0$, where $n$ is a
(finite) positive real number.\footnote{More precisely, $\rmd\ln p/\rmd\ln\rho\to1+1/n$ as $\rho\to0$.}  In this case $\rho$ and $p$ are
increasing functions of $h$ for all $h>0$, and $\rho\propto h^n$,
$p\propto h^{n+1}$ and $v_\rms^2\propto h$ in the limit $h\to0$; the
simplest example of such a barotropic relation is the polytrope, in
which these power laws hold for all $h>0$.  The body then exists
where $h>0$, and can have a free surface on which $h=0$, while the
normal enthalpy gradient is typically non-zero there.  If $n<1$, the
normal density gradient is weakly singular on the surface.

We will also consider the case of a homogeneous incompressible fluid,
for which $\rho=\cst$.  Although this case resembles the limit $n\to0$
of the family of polytropes, it is best treated separately because the
density is non-zero at the free surface.

\subsection{Linearized equations}

We consider a steady axisymmetric body (`body~1') that is weakly
perturbed by a tidal gravitational potential $\Psi$, which need not be
steady or axisymmetric.  The tidal potential is generated by a
external body (`body~2') and satisfies Laplace's equation
$\nabla^2\Psi=0$ within body~1.

The linearized equations governing the response of body~1 are
\begin{equation}
  \ddot\bxi+2\bOmega\times\dot\bxi=-\bnabla W,
\label{motion}
\end{equation}
\begin{equation}
  W=h'+\Phi'+\Psi,
\label{w}
\end{equation}
\begin{equation}
  \rho'=-\bnabla\cdot(\rho\bxi),
\label{mass}
\end{equation}
\begin{equation}
  \nabla^2\Phi'=4\pi G\rho',
\label{poisson}
\end{equation}
where the prime denotes an Eulerian perturbation, the dot denotes
$\p/\p t$, and $\bxi$ is the displacement, such that $\bu'=\dot\bxi$.
Note that $\Phi'$ represents the internal (self-) gravitational
potential perturbation, while $\Psi$ is the external (tidal)
potential.  Outside the fluid, $\Phi'$ satisfies Laplace's equation
and decays to zero as $|\br|\to\infty$.  Equation~(\ref{mass}) implies
that
\begin{equation}
  h'=-v_\rms^2(\bnabla\cdot\bxi)-\bxi\cdot\bnabla h.
\label{h'}
\end{equation}

The desired solution of these equations is such that $\bxi$,
$\bnabla\cdot\bxi$, $W$ and $\bnabla W$ are bounded everywhere within
the fluid, while $\Phi'$ and $\bnabla\Phi'$ are bounded and continuous
everywhere and tend to zero as $|\br|\to\infty$.  If $n<1$, then
$\rho'$ is weakly singular on the surface, but
equation~(\ref{poisson}) still has a solution such that $\Phi'$ and
$\bnabla\Phi'$ are bounded and continuous.

The body may contain a solid core.  Since a discussion of solid
mechanics is beyond the scope of this paper, we assume that the core
is perfectly rigid.  The boundary condition $\bxi\cdot\bn=0$ then
applies on the surface of the core.  However, we will also consider a
fluid core in Section~\ref{s:fluid core}.

\subsection{Love numbers, energy and angular momentum}

From the above linearized equations and boundary conditions, after
some integration by parts, we obtain an energy equation of the form
\begin{equation}
  \f{\rmd E}{\rmd t}=P,
\end{equation}
where 
\begin{equation}
  E=\int\f{1}{2}\rho\left(|\dot\bxi|^2+\f{h'^2}{v_\rms^2}\right)\,\rmd V-\f{1}{8\pi G}\int|\bnabla\Phi'|^2\,\rmd V
\end{equation}
is the (canonical) energy of the perturbation, and
\begin{equation}
  P=\int\rho\dot\bxi\cdot(-\bnabla\Psi)\,\rmd V
\end{equation}
is the power input by the tidal force.
Integrals that involve $\rho$ are carried out over the entire fluid
volume of body~1, while those that do not are carried out over all
space.  The energy $E$ is positive definite for a gravitationally
stable body.

After an integration by parts, we have
\begin{equation}
  P=-\int\Psi\dot\rho'\,\rmd V=-\f{1}{4\pi G}\int\Psi\nabla^2\dot\Phi'\,\rmd V.
\end{equation}
The latter integral can be carried out over any region that includes
the fluid volume of body~1.  Let $V_*$ be any such region that is
simply connected and does not include body~2, so that $\nabla^2\Psi=0$
in $V_*$.  Then, by Green's second identity,
\begin{eqnarray}
  P&=&\f{1}{4\pi G}\int_{V_*}(\dot\Phi'\nabla^2\Psi-\Psi\nabla^2\dot\Phi')\,\rmd V\nonumber\\
  &=&\f{1}{4\pi G}\int_{\p V_*}(\dot\Phi'\bnabla\Psi-\Psi\bnabla\dot\Phi')\cdot\rmd\bS.
\end{eqnarray}

We assume that $V_*$ can be chosen such that its boundary $\p V_*$ is
a sphere of some radius $a$ centred on body~1, i.e.\ that the two
bodies are sufficiently distant that they can be separated by a
spherical boundary.  Then, in the vicinity of $\p V_*$, $\Psi$ can be
represented using an interior multipole expansion in solid spherical
harmonics, while $\Phi'$ can be represented using an exterior
multipole expansion.  Thus
\begin{equation}
  \Psi=\sum_{l,m}\Psi_l^m(t)\left(\f{r}{R}\right)^lY_l^m(\theta,\phi),
\end{equation}
\begin{equation}
  \Phi'=\sum_{l,m}\Phi_l^m(t)\left(\f{r}{R}\right)^{-(l+1)}Y_l^m(\theta,\phi),
\end{equation}
where $R$ is the nominal (e.g.~equatorial) radius of body~1 and the
sum is over integers $l\ge2$ and $-l\le m\le l$.  Since $\Psi$ and
$\Phi'$ are real, we have $\Psi_l^{-m}=\left(\Psi_l^m\right)^*$ and
$\Phi_l^{-m}=\left(\Phi_l^m\right)^*$.  Having chosen $\p V_*$ to be
the sphere of radius $a$, we find
\begin{equation}
  P=\sum_{l,m}\f{(2l+1)R}{4\pi G}\,\dot\Phi_l^m\left(\Psi_l^m\right)^*,
\end{equation}
which is independent of the parameter $a$, as expected.

An initial-value problem can be solved by Fourier (or Laplace)
transform methods.  If the tidal potential is applied for a finite
time or decays sufficiently rapidly as $t\to\infty$, the total amount
of energy transferred to the body is, by Parseval's theorem,
\begin{eqnarray}
  \Delta E&=&\int_{-\infty}^\infty P\,\rmd t\nonumber\\
  &=&\sum_{l,m}\f{(2l+1)R}{4\pi G}\int_{-\infty}^\infty\dot\Phi_l^m(t)[\Psi_l^m(t)]^*\,\rmd t\nonumber\\
  &=&\sum_{l,m}\f{(2l+1)R}{4\pi G}\int_{-\infty}^\infty(-\rmi\omega)\tilde\Phi_l^m(\omega)[\tilde\Psi_l^m(\omega)]^*\,\f{\rmd\omega}{2\pi},\nonumber\\
\end{eqnarray}
where
\begin{equation}
  \tilde\Psi_l^m(\omega)=\int_{-\infty}^\infty\Psi_l^m(t)\,\rme^{\rmi\omega t}\,\rmd t,
\end{equation}
etc., denote temporal Fourier transforms.  Taking the Fourier
transform of the linearized equations (\ref{motion})--(\ref{poisson})
leads to a purely spatial problem in which the frequency $\omega$
(which may in general be treated as a complex number) appears as a
parameter.  For a body of arbitrary shape, the solution of this
problem yields a linear relation of the form
\begin{equation}
  \tilde\Phi_l^m(\omega)=\sum_{l',m'}k_{l,l'}^{m,m'}(\omega)\tilde\Psi_{l'}^{m'}(\omega),
\label{love}
\end{equation}
involving an array of potential Love numbers
$k_{l,l'}^{m,m'}(\omega)$.  The Love numbers are dimensionless complex
coefficients that quantify the frequency-dependent response of the
body to tidal forcing; since gravity is the only means of
communication between the two bodies, all the information relevant to
spin--orbit coupling and tidal evolution is contained within them.
If the body is spherically symmetric then $k_{l,l'}^{m,m'}(\omega)=0$
unless $l'=l$ and $m'=m$.  Let us assume instead that the body is
axisymmetric (in the selected spherical polar coordinate system), so
that $k_{l,l'}^{m,m'}(\omega)=0$ unless $m'=m$.  Since the linearized
equations are invariant under complex conjugation, we also have
$k_{l,l'}^{m,m}(\omega)=[k_{l,l'}^{-m,-m}(-\omega^*)]^*$.  Then
\begin{eqnarray}
  \lefteqn{\Delta E=\sum_{l,m}\sum_{l'}\f{(2l+1)R}{4\pi G}}&\nonumber\\
  &&\times\int_{-\infty}^\infty(-\rmi\omega)k_{l,l'}^{m,m}(\omega)\tilde\Psi_{l'}^m(\omega)[\tilde\Psi_l^m(\omega)]^*\,\f{\rmd\omega}{2\pi}.
\end{eqnarray}

Consider an idealized problem in which only a single component
$\Psi_l^m$ of the tidal potential is applied.  In fact, because $\Psi$
is real, $\Psi_l^{-m}=\left(\Psi_l^m\right)^*$ must also be present,
and their Fourier transforms are related by
$\tilde\Psi_l^{-m}(\omega)=[\tilde\Psi_l^m(-\omega^*)]^*$.  Then, if
$m\ne0$,
\begin{eqnarray}
  \Delta E&=&\f{(2l+1)R}{4\pi G}\int_{-\infty}^\infty(-\rmi\omega)k_l^m(\omega)|\tilde\Psi_l^m(\omega)|^2\,\f{\rmd\omega}{2\pi}\nonumber\\
  &&+\f{(2l+1)R}{4\pi G}\int_{-\infty}^\infty(-\rmi\omega)k_l^{-m}(\omega)|\tilde\Psi_l^{-m}(\omega)|^2\,\f{\rmd\omega}{2\pi}\nonumber\\
  &=&\f{(2l+1)R}{4\pi G}\int_{-\infty}^\infty(-\rmi\omega)k_l^m(\omega)|\tilde\Psi_l^m(\omega)|^2\,\f{\rmd\omega}{2\pi}\nonumber\\
  &&+\f{(2l+1)R}{4\pi G}\int_{-\infty}^\infty(-\rmi\omega)[k_l^m(-\omega)]^*|\tilde\Psi_l^m(-\omega)|^2\,\f{\rmd\omega}{2\pi}\nonumber\\
  &=&\f{(2l+1)R}{4\pi G}\int_{-\infty}^\infty(-\rmi\omega)k_l^m(\omega)|\tilde\Psi_l^m(\omega)|^2\,\f{\rmd\omega}{2\pi}\nonumber\\
  &&+\f{(2l+1)R}{4\pi G}\int_{-\infty}^\infty(+\rmi\omega)[k_l^m(\omega)]^*|\tilde\Psi_l^m(\omega)|^2\,\f{\rmd\omega}{2\pi}\nonumber\\
  &=&2\,\f{(2l+1)R}{4\pi G}\int_{-\infty}^\infty\omega\,\mathrm{Im}[k_l^m(\omega)]|\tilde\Psi_l^m(\omega)|^2\,\f{\rmd\omega}{2\pi},
\label{klm}
\end{eqnarray}
where $k_l^m(\omega)=k_{l,l}^{m,m}(\omega)$ is the potential Love
number that quantifies the part of the response that has the same form
as the applied tidal potential.  (If instead $m=0$, then the factor of
$2$ is absent from this result.)  Similarly, if $J$ represents the
component of angular momentum parallel to the axis of the coordinate
system, then the quantity transferred is
\begin{equation}
  \Delta J=2\,\f{(2l+1)R}{4\pi G}\int_{-\infty}^\infty m\,\mathrm{Im}[k_l^m(\omega)]|\tilde\Psi_l^m(\omega)|^2\,\f{\rmd\omega}{2\pi}.
\end{equation}

If the body is stable, the Love number should be an analytic function
of $\omega$ in the upper complex half-plane.  In the absence of
dissipative processes it may have singularities on the real axis,
indicating an unbounded resonant response.

The Fourier-transformed problem is equivalent to one in which the
forcing and the response are assumed to depend harmonically on time
through a common factor $\rme^{-\rmi\omega t}$.  If $\omega$ is real,
and dissipative terms are included in the equations to move the
singularities below the real axis, then there is a steady rate of
energy transfer, equal to the rate of energy dissipation and
proportional to $\omega\,\mathrm{Im}[k_l^m(\omega)]$.  Specifically,
if $\Psi=\mathrm{Re}[A(r/R)^lY_l^m(\theta,\phi)\,\rme^{-\rmi\omega
  t}]$, where $A$ is an arbitrary constant of appropriate dimensions,
then this rate is
\begin{equation}
  P=\f{(2l+1)R}{4\pi G}\f{1}{2}|A|^2\omega\,\mathrm{Im}[k_l^m(\omega)],
\label{rate}
\end{equation}
and the tidal torque on the body differs by a factor of $m/\omega$.
(In the case $m=0$ this expression for $P$ represents a time-average.)

The rate of energy transfer is a frame-dependent quantity.  If the
problem is analysed in a frame in which the body rotates, then the
rate of energy transfer differs from the dissipation rate.  The
difference is equal to the rate of change of rotational energy of the
body under the action of the tidal torque.  For a similar reason, in a
differentially rotating body there is no simple global relation
between the dissipation rate and the rates of energy and angular
momentum transfer.

\subsection{Low-frequency oscillations in a slowly rotating body}
\label{s:asymptotics}

A body of the type considered in Section~\ref{s:equilibrium} is
capable of several types of oscillation: acoustic waves, surface
gravity waves and inertial waves.  The frequencies of the first two
types of oscillation are typically greater than the characteristic
dynamical frequency $(GM/R^3)^{1/2}$ of the body, while the
frequencies of the third type of oscillation are typically comparable
to $\Omega$.  We define the dimensionless parameter $\epsilon$ via
\begin{equation}
  \Omega=\epsilon\left(\f{GM}{R^3}\right)^{1/2}.
\label{epsilon}
\end{equation}
When the body is slowly rotating ($\epsilon^2\ll1$) and the
equilibrium structure is close to spherical symmetry, the inertial
waves are well separated in frequency from the acoustic and surface
gravity waves.  The frequencies of tidal oscillations typically lie in
the range of inertial waves.

An asymptotic theory of low-frequency forced oscillations in a slowly
rotating body was developed by \citet{2004ApJ...610..477O}.  We
further develop this theory here for the case of barotropic fluids,
but without using a formal asymptotic expansion; similar
approximations have been used by
\citet{2005ApJ...635..674W,2005ApJ...635..688W} and
\citet{2005MNRAS.364L..66P}.  We are interested in a low-frequency
limit in which the tidal frequency is comparable to $\Omega$ and
therefore $O(\epsilon)$, with $\epsilon\ll1$, while the dynamical
frequency\footnote{In effect, we are considering a family of slowly
  rotating bodies labelled by the small parameter $\epsilon$, and
  adopting a system of units such that $M$ and $R$ are the same for
  each body, while $\Omega\propto\epsilon$.} is $O(1)$.  Assuming that
the tidal potential $\Psi=O(1)$ (which is a convenient but arbitrary
choice for a linear analysis), we find that the scalings $\bxi=O(1)$,
$W=O(\epsilon^2)$, $h'=O(1)$, $\Phi'=O(1)$ are required in order to
satisfy the linearized equations.  To leading order, therefore, we
wish to solve the system
\begin{equation}
  \ddot\bxi+2\bOmega\times\dot\bxi=-\bnabla W,
\label{motion2}
\end{equation}
\begin{equation}
  h'+\Phi'+\Psi=0,
\label{w2}
\end{equation}
\begin{equation}
  \rho'=-\bnabla\cdot(\rho\bxi),
\label{mass2}
\end{equation}
\begin{equation}
  \nabla^2\Phi'=4\pi G\rho',
\label{poisson2}
\end{equation}
in which the $W$ term has dropped out of equation~(\ref{w2}).

Furthermore, to this level of approximation, the basic state can be
assumed to be spherically symmetric and is conveniently described
using spherical polar coordinates $(r,\theta,\phi)$, with the
coordinate axis coinciding with the axis of rotation.  The fluid occupies
the region $\alpha R<r<R$, if we assume the solid core to be
spherical
and of fractional radius $\alpha$ with $0\le\alpha<1$, and is in
hydrostatic equilibrium with $\rmd h/\rmd r=-g$, where $g(r)$ is the
inward gravitational acceleration.  The boundary conditions are as
described above.  Since the regular solution satisfies $h'=g\xi_r$ at
$r=R$ (see equation~\ref{h'}), we have
\begin{equation}
  \xi_r=0\qquad\hbox{at}\quad r=\alpha R,
\end{equation}
\begin{equation}
  \xi_r=-\f{\Phi'+\Psi}{g}\qquad\hbox{at}\quad r=R.
\end{equation}

Combining equations~(\ref{w2}) and~(\ref{poisson2}), we obtain a type
of inhomogeneous Helmholtz equation \citep[cf.][]{1966AnAp...29..313Z},
\begin{equation}
  \nabla^2\Phi'+\f{4\pi G\rho}{v_\rms^2}(\Phi'+\Psi)=0,
\label{helmholtz}
\end{equation}
for the potential perturbation within the fluid, which simplifies to
Laplace's equation in the surrounding vacuum and in the solid core.
For given $\Psi$ of an appropriate form, this equation has a unique
solution such that $\Phi'$ and $\bnabla\Phi'$ are bounded and
continuous everywhere and tend to zero as $|\br|\to\infty$.

In particular, the unforced problem $\Psi=0$ has the solution
$\Phi'=0$, which requires $\rho'=h'=0$, as well as $\xi_r=0$ at $r=R$.
Free oscillations in the low-frequency limit, which correspond to
unforced inertial waves, therefore satisfy the anelastic constraint
$\bnabla\cdot(\rho\bxi)=0$ and the rigid boundary conditions $\xi_r=0$
on surfaces with both vacuum and solid.  These conditions ensure that
the otherwise unknown $W$ in equation~(\ref{motion2}) can be
determined instantaneously in terms of $\dot\bxi$, because we obtain a
modified Poisson equation $\bnabla\cdot(\rho\bnabla
W)=-\bnabla\cdot(2\rho\bOmega\times\dot\bxi)$ with Neumann boundary
conditions (or regularity at $r=R$), whose solution is unique except
for an irrelevant additive constant (or function of time).

In the forced problem $\Psi\ne0$, we must have $\Phi'\ne0$,
$\rho'\ne0$ and $h'\ne0$.  It is natural to separate the forced
solution into \textit{non-wavelike} and \textit{wavelike} parts:
$\bxi=\bxi_\rmnw+\bxi_\rmw$, etc.  The rationale behind this
decomposition is that the non-wavelike part is an instantaneous
hydrostatic response to the tidal potential; this is calculated by
neglecting the Coriolis force, which is responsible for the existence
of inertial waves, but it involves the correct Eulerian perturbations
$\Phi'$, $\rho'$ and $h'$ to satisfy the tidal forcing.  The wavelike
part merely corrects for the neglect of the Coriolis force.  The two
parts therefore satisfy the following equations, which are a natural
decomposition of equations~(\ref{motion2})--(\ref{poisson2}):
\begin{equation}
  \ddot\bxi_\rmnw=-\bnabla W_\rmnw,
\label{motion3}
\end{equation}
\begin{equation}
  h'_\rmnw+\Phi'_\rmnw+\Psi=0,
\label{w3}
\end{equation}
\begin{equation}
  \rho'_\rmnw=-\bnabla\cdot(\rho\bxi_\rmnw),
\label{mass3}
\end{equation}
\begin{equation}
  \nabla^2\Phi'_\rmnw=4\pi G\rho'_\rmnw,
\label{poisson3}
\end{equation}
for the non-wavelike part, together with
\begin{equation}
  \ddot\bxi_\rmw+2\bOmega\times\dot\bxi_\rmw=-\bnabla W_\rmw+\bff,
\label{motion4}
\end{equation}
\begin{equation}
  \bnabla\cdot(\rho\bxi_\rmw)=0,
\label{mass4}
\end{equation}
for the wavelike part with $\rho'_\rmw=h'_\rmw=\Phi'_\rmw=0$, where
\begin{equation}
  \bff=-2\bOmega\times\dot\bxi_\rmnw
\label{f}
\end{equation}
appears as an effective force per unit mass driving the wavelike part
of the solution.  Since $\rho'_\rmw=h'_\rmw=\Phi'_\rmw=0$, we can omit
the subscript on $\rho'_\rmnw$, $h'_\rmnw$ and $\Phi'_\rmnw$.

The boundary conditions are also naturally decomposed as
\begin{equation}
  \xi_{\rmnw,r}=0\qquad\hbox{at}\quad r=\alpha R,
\label{bc1}
\end{equation}
\begin{equation}
  \xi_{\rmnw,r}=-\f{\Phi'+\Psi}{g}\qquad\hbox{at}\quad r=R,
\label{bc2}
\end{equation}
\begin{equation}
  \xi_{\rmw,r}=0\qquad\hbox{at}\quad r=\alpha R,
\label{bc3}
\end{equation}
\begin{equation}
  \xi_{\rmw,r}=0\qquad\hbox{at}\quad r=R.
\label{bc4}
\end{equation}
This implies that the
wavelike part is driven only by the unbalanced Coriolis force acting
on the non-wavelike part, and not through an inhomogeneous boundary
condition.

Although equations~(\ref{motion3})--(\ref{poisson3}) contain
time-derivatives, they do in fact imply that $\bxi_\rmnw$ is (or may
be assumed to be) \textit{instantaneously} related to the tidal
potential $\Psi$.  This is most easily seen by rewriting
equation~(\ref{motion3}) as $\bxi_\rmnw=-\bnabla X$, where $\ddot
X=W_\rmnw$.  The equations and boundary conditions defining
$\bxi_\rmnw$ are then instantaneous in time.  Note that
\begin{equation}
  \bnabla\cdot(\rho\bnabla X)=\rho'=-\f{\rho}{v_\rms^2}(\Phi'+\Psi).
\label{x}
\end{equation}
The right-hand side of this equation is known from the solution of the
Helmholtz-like equation~(\ref{helmholtz}).  This is, again, a well
posed elliptic boundary-value problem which determines $X$ up to an
additive constant.  The boundary conditions (\ref{bc1}) and
(\ref{bc2}) correspond to $\p X/\p r=0$ at $r=\alpha R$ and regularity
of $X$ at $r=R$.

This procedure does not provide the most general solution of
equations~(\ref{motion3})--(\ref{poisson3}), because the displacement
could also contain a vortical part (not satisfying $\bxi=-\bnabla X$)
with a linear dependence on time, which would not contribute to
$\ddot\bxi$.  However, any such residual displacement, which would
depend on the initial conditions, can be considered to be part of the
wavelike part of the solution.  In other words, the initial conditions
are decomposed such that $\bxi_\rmnw$ is defined as above and does not
depend on the initial data, while $\bxi_\rmw$ contains the two
functional degrees of freedom arising from the choice of initial
values of $\bxi$ and $\dot\bxi$.

The non-wavelike part as defined above is not equivalent to the
conventional `equilibrium tide' as defined by
\citet{1966AnAp...29..313Z} and \citet{1989ApJ...342.1079G}.  The
equilibrium tide is a spheroidal displacement of the form
\begin{equation}
  \bxi_\rme=\be_r\,R_\rme-\be_r\times(\be_r\times\bnabla S_\rme),
\end{equation}
with
\begin{equation}
  R_\rme=-\f{\Phi'+\Psi}{g},
\end{equation}
\begin{equation}
  0=\bnabla\cdot\bxi_\rme=\f{1}{r^2}\f{\p}{\p r}(r^2R_\rme)+\nabla_\rmh^2S_\rme.
\end{equation}
It is supposed to represent the low-frequency limit of the tidal
response, but does not apply in a barotropic body where there is no
stable stratification, as pointed out by \citet{1998ApJ...502..788T}
and \citet{1998ApJ...507..938G}.  The equilibrium tide satisfies all
equations~(\ref{motion3})--(\ref{poisson3}) except the first, because
it is not irrotational in general.  In fact,
\begin{equation}
  \bnabla\times\bxi_\rme=-\be_r\times\bnabla_\rmh\left(R_\rme-\f{\p S_\rme}{\p r}\right),
\end{equation}
and
\begin{equation}
  r\nabla_\rmh^2\left(R_\rme-\f{\p S_\rme}{\p r}\right)=\nabla^2(rR_\rme)=-\nabla^2\left[\f{r}{g}(\Phi'+\Psi)\right],
\end{equation}
which does vanish in a homogeneous body, in which $g\propto r$ and
$\nabla^2\Phi'=0$, but not generally.  In addition, if the body
contains a (perfectly rigid) solid core, the equilibrium tide does not
satisfy the appropriate rigid boundary condition there.

The equations governing the wavelike part of the solution are well
posed as an initial-value problem, even in the absence of dissipative
processes.  In contrast, when the forcing is harmonic in time and the
solution is assumed to have the same periodicity, the equations
governing its spatial structure are hyperbolic and not well posed in
the absence of dissipation for frequencies in the range
$-2\Omega<\omega<2\Omega$ of inertial waves, and singularities may
occur.

The energy equation for the wavelike part is
\begin{equation}
  \f{\rmd}{\rmd t}\int\f{1}{2}\rho|\dot\bxi_\rmw|^2\,\rmd V=\int\rho\dot\bxi_\rmw\cdot\bff\,\rmd V,
\end{equation}
where the integrals are over the fluid volume.  The right-hand side
represents the power input to the wavelike part from the effective
force; the energy comes ultimately from the source of the external
potential.  Only kinetic energy is relevant here because of the
absence of stable stratification and the inability of inertial waves
to elevate the free surface in the low-frequency limit.  If we
introduce viscous or frictional forces to provide a mechanism of
dissipation and to resolve the singularities in the response, then a
dissipative term should be included in this energy equation.

\subsection{Fourier transform and harmonic forcing}

The initial-value problem for the wavelike part can be analysed by
taking a Fourier (or Laplace) transform in time, leading us to
consider the problem
\begin{equation}
  -\rmi\omega\tilde\bu_\rmw+2\bOmega\times\tilde\bu_\rmw=-\bnabla\tilde W_\rmw+\tilde\bff,
\label{ft}
\end{equation}
\begin{equation}
  \bnabla\cdot(\rho\tilde\bu_\rmw)=0,
\end{equation}
where $\tilde\bu_\rmw=-\rmi\omega\tilde\bxi_\rmw$ and
\begin{equation}
  \tilde\bxi_\rmw(\br,\omega)=\int_{-\infty}^\infty\bxi_\rmw(\br,t)\,\rme^{\rmi\omega t}\,\rmd t,
\end{equation}
etc.  The rigid boundary
conditions $\tilde\xi_{\rmw,r}=0$ apply
and we may wish to consider complex frequencies $\omega$ in order to
carry out the inverse transform and to avoid singularities.  The
Fourier-transformed variables $\tilde\bu_\rmw$, etc., are also complex
in general.

If the force is applied for a finite time or decays sufficiently
rapidly as $t\to\infty$, the total amount of energy transferred to the
wavelike part is, by Parseval's theorem,
\begin{eqnarray}
  \Delta E&=&\int_{-\infty}^\infty\int\rho\bu_\rmw\cdot\bff\,\rmd V\,\rmd t\nonumber\\
  &=&\int_{-\infty}^\infty\int\rho\tilde\bu_\rmw\cdot\tilde\bff^*\,\rmd V\,\f{\rmd\omega}{2\pi}.
\end{eqnarray}

The Fourier-transformed problem is formally equivalent to one in which
the forcing  and the  response are assumed  to depend  harmonically on
time  through a  common  factor $\rme^{-\rmi\omega t}$, with  $\omega$
being  complex in  general.  In  this case  the interpretation  of the
equations  is   different,  however.   The  force   and  the  wavelike
displacement  are $\mathrm{Re}[\tilde\bff(\br)\,\rme^{-\rmi\omega t}]$
and     $\mathrm{Re}[\tilde\bxi_\rmw(\br)\,\rme^{-\rmi\omega     t}]$,
respectively, and (when  $\omega$ is real) the average  rate of energy
transfer is
\begin{equation}
  P=\f{1}{2}\,\mathrm{Re}\int\rho\tilde\bu_\rmw\cdot\tilde\bff^*\,\rmd V.
\label{power}
\end{equation}
As noted above, this problem is generally ill posed for real
frequencies in the range of inertial waves, so we may choose to add a
viscous or frictional force to ensure a regular solution.  In that
case the dissipation rate equals the power input for real frequencies.
For example, if the frictional term $-\gamma\tilde\bu_\rmw$ is added
to the right-hand side of equation~(\ref{ft}), where $\gamma>0$ is a
frictional damping rate, then we obtain
\begin{equation}
  \f{1}{2}[\mathrm{Im}(\omega)+\gamma]\int\rho|\tilde\bu_\rmw|^2\,\rmd V=\f{1}{2}\,\mathrm{Re}\int\rho\tilde\bu_\rmw\cdot\tilde\bff^*\,\rmd V.
\end{equation}

We quantify the solution of the harmonic forcing problem as follows.
If the force derives from a tidal potential
$\Psi=\mathrm{Re}[A(r/R)^lY_l^m(\theta,\phi)\,\rme^{-\rmi\omega t}]$,
where $A$ is an arbitrary constant of appropriate dimensions, we let
\begin{equation}
  \int\rho\tilde\bxi_\rmw\cdot\tilde\bff^*\,\rmd V=\f{(2l+1)R}{4\pi G}|A|^2K_l^m(\omega),
\label{K}
\end{equation}
thereby defining the complex dimensionless response function
$K_l^m(\omega)$, which is linearly related to the response of the
fluid and has the Hermitian symmetry property
$K_l^m(\omega)=[K_l^{-m}(-\omega^*)]^*$.  This is defined by analogy
with equation~(\ref{rate}), which relates the energy transfer to the
potential Love number $k_l^m(\omega)$.  Comparing
equations~(\ref{power}) and~(\ref{K}), we find that, for real frequencies,
\begin{equation}
  P=\f{(2l+1)R}{4\pi G}\f{1}{2}|A|^2\omega\,\mathrm{Im}[K_l^m(\omega)],
\end{equation}
which agrees with equation~(\ref{rate}) if
$\mathrm{Im}[k_l^m(\omega)]=\mathrm{Im}[K_l^m(\omega)]$.  Indeed, in
the low-frequency limit considered here, we expect that
$k_l^m(\omega)$ is real to a first approximation, and that the
dissipative part of the response is given by
$\mathrm{Im}[k_l^m(\omega)]=\mathrm{Im}[K_l^m(\omega)]=O(\epsilon^2)$.
Since the effective force driving the wavelike tide is a Coriolis
force, $\bff\propto\Omega$, the response $\bxi_\rmw$ is also
proportional to $\Omega$, and therefore $K_l^m\propto\Omega^2$.  The
reason for quantifying the response using equation~(\ref{K}) rather
than equation~(\ref{love}) is that, in the low-frequency limit, the
gravitational potential perturbation $\Phi'$ is a quantity that is not
readily available, hidden as it is within $W$.

When a single potential component $\Psi_l^m(t)$ (having a general
time-dependence that decays sufficiently at large $t$, and being
accompanied by $\Psi_l^{-m}$ when $m\ne0$) is applied, the total
energy transferred to the wavelike tide is, by analogy with
equation~(\ref{klm}),
\begin{equation}
  \Delta E=2\,\f{(2l+1)R}{4\pi G}\int_{-\infty}^\infty\omega\,\mathrm{Im}[K_l^m(\omega)]|\tilde\Psi_l^m(\omega)|^2\,\f{\rmd\omega}{2\pi},
\label{deltae}
\end{equation}
or half this quantity in the case $m=0$.

\subsection{Impulsive forcing}
\label{s:impulse}

We now consider a special tidal forcing problem in which the wavelike
part experiences an impulsive effective force of the form
\begin{equation}
  \bff=\hat\bff(\br)\,\delta(t),
\end{equation}
where $\delta(t)$ is the Dirac delta function.  Since $\bff$ is related
to $\dot\bxi_\rmnw$ through equation~(\ref{f}) and $\bxi_\rmnw$ is
instantaneously related to the tidal potential, this means that the
potential should be of the form
\begin{equation}
  \Psi=\hat\Psi(\br)H(t),
\end{equation}
where $H(t)$ is the Heaviside step function.  This, rather than the
more obvious $\Psi\propto\delta(t)$, is the appropriate type of
impulse problem to consider within the low-frequency approximation.
The impulse is supposed to occur slowly enough that the anelastic
constraint holds, i.e.\ slowly compared to the sound crossing time,
but fast compared to the rotation period.  The idea is to excite a
broad spectrum of inertial waves but no surface gravity or acoustic
waves.  \citet{2010MNRAS.407.1631P} in their Appendix~B1 discuss a
problem of this type, but without explicit calculations; they do,
however, solve initial-value problems in which inertial waves are
excited by a parabolic tidal encounter.

Assuming that the fluid is at rest before the impulse, immediately
afterwards it will have a wavelike velocity given by
\begin{equation}
  \hat\bu_\rmw=\hat\bff-\bnabla\hat W_\rmw,
\label{uhat}
\end{equation}
where $\hat W_\rmw$ is arranged to satisfy the anelastic constraint
\begin{equation}
  \bnabla\cdot(\rho\hat\bu_\rmw)=0
\end{equation}
and the boundary conditions $\hat u_{\rmw,r}=0$.  Note that
$\hat\bu_\rmw$ is just the initial velocity immediately after the
impulse; it will subsequently oscillate as a collection of inertial
waves and be damped, if dissipative terms are included in the
equations.  Equation~(\ref{uhat}) is derived by integrating
equation~(\ref{motion4}) over an arbitrarily small time-interval that
includes the instant $t=0$.  Note that the Coriolis force acting on
the wavelike velocity has no effect during the impulse process; the
only role of the Coriolis force is to provide the effective force
$\bff$ from its action on the non-wavelike velocity.

We now require a procedure to calculate the impulsive response and the
associated energy transfer
\begin{equation}
  \hat E=\f{1}{2}\int\rho|\hat\bu_\rmw|^2\,\rmd V.
\end{equation}
We consider a tidal potential with the spatial structure
\begin{equation}
  \hat\Psi=\mathrm{Re}\left[\hat\Psi_l(r)Y_l^m(\theta,\phi)\right]=\mathrm{Re}\left[A\left(\f{r}{R}\right)^lY_l^m(\theta,\phi)\right].
\end{equation}
The associated internal gravitational potential perturbation [also
proportional to $H(t)$],
\begin{equation}
  \hat\Phi'=\mathrm{Re}\left[\hat\Phi'_l(r)Y_l^m(\theta,\phi)\right],
\end{equation}
satisfies the Helmholtz-like equation~(\ref{helmholtz}), which reduces
to the ordinary differential equation (ODE)
\begin{equation}
  \f{1}{r^2}\f{\rmd}{\rmd r}\left(r^2\f{\rmd\hat\Phi'_l}{\rmd r}\right)-\f{l(l+1)}{r^2}\hat\Phi'_l+\f{4\pi G\rho}{v_\rms^2}(\hat\Phi'_l+\hat\Psi_l)=0
\label{helmholtzode}
\end{equation}
in $\alpha R<r<R$.  Matching to decaying solutions of Laplace's
equation in the solid core and the exterior vacuum provides the
boundary conditions $\rmd\hat\Phi'_l/\rmd r=l\hat\Phi'_l/r$ at
$r=\alpha R$ and $\rmd\hat\Phi'_l/\rmd r=-(l+1)\hat\Phi'_l/r$ at
$r=R$.  The associated non-wavelike tide [also proportional to $H(t)$]
is
\begin{equation}
  \hat\bxi_\rmnw=-\bnabla\hat X,
\end{equation}
where $\hat X=\mathrm{Re}[\hat X_l(r)Y_l^m(\theta,\phi)]$ satisfies
equation~(\ref{x}), which reduces to the ODE
\begin{equation}
  \f{1}{r^2}\f{\rmd}{\rmd r}\left(r^2\rho\f{\rmd\hat X_l}{\rmd r}\right)-\f{l(l+1)}{r^2}\rho\hat X_l=-\f{\rho}{v_\rms^2}(\hat\Phi'_l+\hat\Psi_l)
\label{xode}
\end{equation}
in $\alpha R<r<R$, together with $\rmd\hat X_l/\rmd r=0$ at $r=\alpha
R$ and regularity of $\hat X_l$ at $r=R$.  From $\hat\bxi_\rmnw$ we
deduce the impulsive effective force
$\hat\bff=-2\bOmega\times\hat\bxi_\rmnw$, which has both spheroidal
and toroidal parts.  Using standard methods of projection on to vector
spherical harmonics \citep[e.g.][]{2004ApJ...610..477O} we find from
equation~(\ref{uhat}) that the impulsive wavelike velocity is of the
form
\begin{eqnarray}
  \lefteqn{\hat\bu_\rmw=\mathrm{Re}\left[\be_r\,\hat a_l(r)Y_l^m+r^2\hat b_l(r)\bnabla Y_l^m\right.}&\nonumber\\
  &&\left.-r^2\hat c_{l-1}(r)\,\be_r\times\bnabla Y_{l-1}^m-r^2\hat c_{l+1}(r)\,\be_r\times\bnabla Y_{l+1}^m\right],
\end{eqnarray}
where $\hat a_l$ and $\hat b_l$ represent the spheroidal part and $\hat c_{l\pm1}$ the toroidal part.  These components satisfy
\begin{equation}
  \hat a_l=-\f{2\rmi m\Omega\hat X_l}{r}-\f{\rmd\hat W_l}{\rmd r},
\end{equation}
\begin{equation}
  \hat b_l=-\f{2\rmi m\Omega}{l(l+1)r^2}\left(r\f{\rmd\hat X_l}{\rmd r}+\hat X_l\right)-\f{\hat W_l}{r^2},
\end{equation}
\begin{equation}
  \hat c_{l-1}=-\f{2\Omega\tilde q_l}{r^2}\left[r\f{\rmd\hat X_l}{\rmd r}+(l+1)\hat X_l\right],
\end{equation}
\begin{equation}
  \hat c_{l+1}=\f{2\Omega\tilde q_{l+1}}{r^2}\left(r\f{\rmd\hat X_l}{\rmd r}-l\hat X_l\right),
\end{equation}
\begin{equation}
  \f{1}{r^2}\f{\rmd}{\rmd r}(r^2\rho\hat a_l)-l(l+1)\rho\hat b_l=0,
\end{equation}
where $\hat W_\rmw=\mathrm{Re}[\hat W_l(r)Y_l^m(\theta,\phi)]$ and
\begin{equation}
  \tilde q_n=\f{1}{n}\left(\f{n^2-m^2}{4n^2-1}\right)^{1/2}
\end{equation}
is a coefficient arising from the coupling of spheroidal and toroidal
velocity components by the Coriolis force.  Therefore $\hat W_l$
satisfies
\begin{equation}
  \f{1}{r^2}\f{\rmd}{\rmd r}\left(r^2\rho\f{\rmd\hat W_l}{\rmd r}\right)-\f{l(l+1)}{r^2}\rho\hat W_l=-\f{2\rmi m\Omega}{r}\f{\rmd\rho}{\rmd r}\hat X_l,
\label{wode}
\end{equation}
with boundary conditions $\rmd\hat W_l/\rmd r=-2\rmi m\Omega\hat X_l/r$ (corresponding to $\hat a_l=0$) at $r=\alpha R$ and $r=R$.

The impulsive energy transfer is then found to be
\begin{equation}
  \hat E=\hat E_l+\hat E_{l-1}+\hat E_{l+1},
\end{equation}
with
\begin{equation}
  \hat E_l=\f{1}{4}\int_{\alpha R}^R\rho r^2\left[|\hat a_l|^2+l(l+1)r^2|\hat b_l|^2\right]\,\rmd r,
\end{equation}
\begin{equation}
  \hat E_{l-1}=\f{1}{4}\int_{\alpha R}^R\rho r^2\left[l(l-1)r^2|\hat c_{l-1}|^2\right]\,\rmd r,
\end{equation}
\begin{equation}
  \hat E_{l+1}=\f{1}{4}\int_{\alpha R}^R\rho r^2\left[(l+1)(l+2)r^2|\hat c_{l+1}|^2\right]\,\rmd r,
\end{equation}
although these expressions should be doubled in the case $m=0$,
assuming that $A$ is real.  Let us now compare this result with
equation~(\ref{deltae}).  The impulsive (switched-on) potential
corresponds to\footnote{It may be better to think of the impulse as
  $H(t)\,\rme^{-\epsilon t}$ with $\epsilon\searrow0$.  Then
  $\tilde\Psi_l^m(\omega)=\rmi A/[2(\omega+\rmi\epsilon)]$, again with
  $\epsilon\searrow0$.}
\begin{equation}
  \tilde\Psi_l^m(\omega)=\f{\rmi A}{2\omega},
\end{equation}
and so
\begin{equation}
  \hat E=2\,\f{(2l+1)R}{4\pi G}\f{|A|^2}{8\pi}\int_{-\infty}^\infty\mathrm{Im}[K_l^m(\omega)]\,\f{\rmd\omega}{\omega}.
\end{equation}

\subsection{Interpretation}

The previous subsection shows that the energy transferred to the
wavelike tide in the impulse problem considered above corresponds to
a certain frequency-average of the imaginary part of the Love number
of the body.  Although the integral as written is over all values of
$\omega$, in fact we need integrate only over the range
$-2\Omega<\omega<2\Omega$ where inertial waves occur, and where the
low-frequency approximation holds.\footnote{Note that, if a
  low-frequency approximation is not made, and an impulsive
  gravitational force arising from a tidal potential
  $\propto\delta(t)$ is applied, then the energy transfer is instead
  proportional to
  $\int\mathrm{Im}[k_l^m(\omega)]\,\omega\,\rmd\omega$.  This integral
  is expected to be dominated by surface-gravity modes of higher
  frequency.}  If no dissipation is included in the equations,
$K_l^m(\omega)$ has singularities on the real $\omega$ axis and the
integral can be computed only by deforming the integration contour
around them in an appropriate way.  In that case the integral
describes the energy given to waves that are never damped.  If,
however, some dissipation is included, then the waves are ultimately
damped and the integral describes the energy dissipated as well as
that transferred.

We see, therefore, that the dimensionless quantity
$\int_{-\infty}^\infty\mathrm{Im}[K_l^m(\omega)]\,\rmd\omega/\omega$,
which is a frequency-averaged measure of the dissipative properties of
the body in the low-frequency range corresponding to inertial waves,
can be directly related to the energy transferred in an impulsive
interaction.  Furthermore, this quantity can be computed directly by
solving a simple set of ordinary differential equations.  This
procedure is much easier than computing an inertial wave with a
harmonic time-dependence, because during the impulsive interaction the
Coriolis force has no time to act on the wavelike velocity; the only
role of the Coriolis force is to provide the effective force from its
action on the non-wavelike velocity.  The solution depends on the
internal structure of the body but not on its dissipative properties;
this solution is smooth and free from boundary layers.  Following the
impulse, this velocity field will resolve into a collection of
inertial waves and be damped, if dissipative terms are included in the
equations, but for our present purposes we need not consider this
subsequent evolution.  In the following subsections we compute the
quantity
$\int_{-\infty}^\infty\mathrm{Im}[K_l^m(\omega)]\,\rmd\omega/\omega$
for some simple interior models and compare with numerical solutions
of the frequency-dependent response functions of such bodies.

\subsection{Application to a homogeneous fluid}
\label{s:homogeneous}

We first consider, as in \citet{2009MNRAS.396..794O}, a body
consisting of a homogeneous incompressible fluid of density $\rho$,
possibly containing a perfectly rigid solid core of fractional radius
$\alpha$.  The mean density of the body is $\bar\rho$.

In this case the Helmholtz-like equation~(\ref{helmholtzode}) becomes
\begin{eqnarray}
  \lefteqn{\f{1}{r^2}\f{\rmd}{\rmd r}\left(r^2\f{\rmd\hat\Phi'_l}{\rmd r}\right)-\f{l(l+1)}{r^2}\hat\Phi'_l}&\nonumber\\
  &&+\f{4\pi G\rho}{g}(\hat\Phi'_l+\hat\Psi_l)\,\delta(r-R)=0,
\end{eqnarray}
where $g=GM/R^2=4\pi G\bar\rho R/3$ is the surface gravity.  Note
that, in this incompressible model, the sound speed is infinite, and
the Eulerian density perturbation is given by
\begin{equation}
  \rho'=-\bxi\cdot\bnabla\rho=\xi_r\rho\,\delta(r-R).
\end{equation}
Given $\hat\Psi_l=A(r/R)^l$, the solution is $\hat\Phi'_l=B(r/R)^l$
for $0\le r\le R$ and $\hat\Phi'_l=B(R/r)^{l+1}$ for $r\ge R$, where
$(2l+1)B=(3\rho/\bar\rho)(A+B)$.

Since $\bnabla\cdot\bxi=0$ for an incompressible fluid, $\hat X$
satisfies Laplace's equation and the solution satisfying the boundary
conditions~(\ref{bc1}) and~(\ref{bc2}) is
\begin{equation}
  \hat X_l=C\left[\left(\f{r}{R}\right)^l+\left(\f{l}{l+1}\right)\alpha^{2l+1}\left(\f{R}{r}\right)^{(l+1)}\right],
\end{equation}
with
\begin{equation}
  C=\f{(A+B)R}{l(1-\alpha^{2l+1})g}.
\end{equation}

$\hat W_\rmw$ also satisfies Laplace's equation and the relevant
solution is
\begin{equation}
  \hat W_l=-\f{2\rmi m\Omega C}{l}\left[\left(\f{r}{R}\right)^l-\left(\f{l}{l+1}\right)^2\alpha^{2l+1}\left(\f{R}{r}\right)^{(l+1)}\right].
\end{equation}

Thus
\begin{equation}
  \hat a_l=\hat b_l=0,
\end{equation}
\begin{equation}
  \hat c_{l-1}=-\f{2\Omega\tilde q_l}{r^2}C(2l+1)\left(\f{r}{R}\right)^l,
\end{equation}
\begin{equation}
  \hat c_{l+1}=-\f{2\Omega\tilde q_{l+1}}{r^2}C(2l+1)\left(\f{l}{l+1}\right)\alpha^{2l+1}\left(\f{R}{r}\right)^{(l+1)},
\end{equation}
and so
\begin{equation}
  \hat E_l=0,
\end{equation}
\begin{equation}
  \hat E_{l-1}=\rho\Omega^2\tilde q_l^2|C|^2l(l-1)(2l+1)(1-\alpha^{2l+1})R,
\end{equation}
\begin{equation}
  \hat E_{l+1}=\rho\Omega^2\tilde q_{l+1}^2|C|^2l^2\left(\f{l+2}{l+1}\right)(2l+1)\alpha^{2l+1}(1-\alpha^{2l+1})R.
\end{equation}
The total impulse energy $\hat E$ is the sum of these three expressions.
Now $\tilde q_l^2\propto(l^2-m^2)$ vanishes for sectoral harmonics
($|m|=l$), while $\tilde q_{l+1}^2$ is non-zero.  For sectoral
harmonics, therefore, the impulse energy has a strong dependence on
core size, being proportional to $\alpha^{2l+1}$ for small $\alpha$.
For tesseral harmonics ($|m|<l$), however, $\hat E$ remains
significant in the limit $\alpha\to0$ because of the contribution of
$\hat E_{l-1}$.  We also see that the impulse energy is proportional
to $\Omega^2$, as expected.

\begin{figure*}
\centerline{\epsfbox{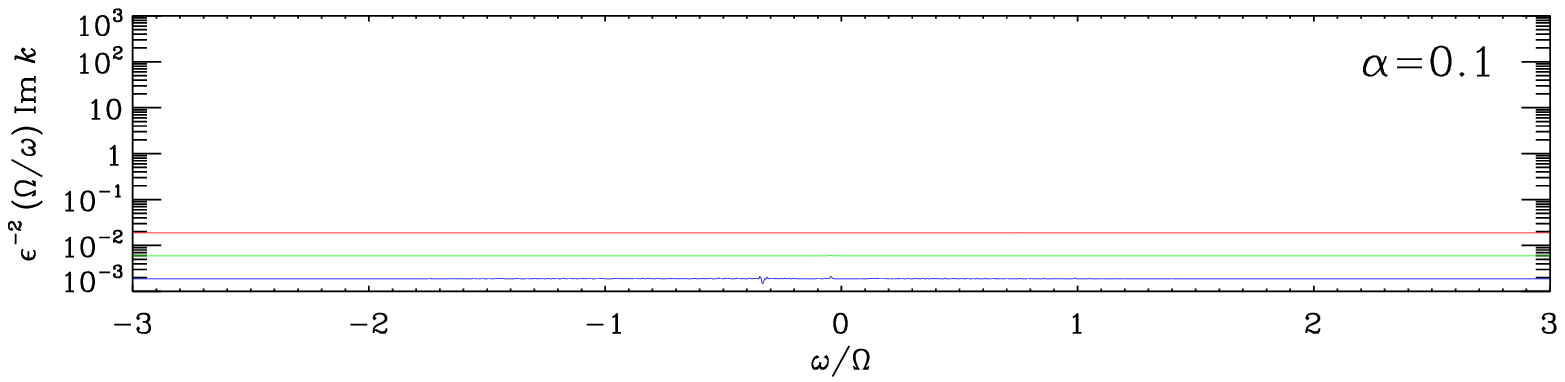}}
\centerline{\epsfbox{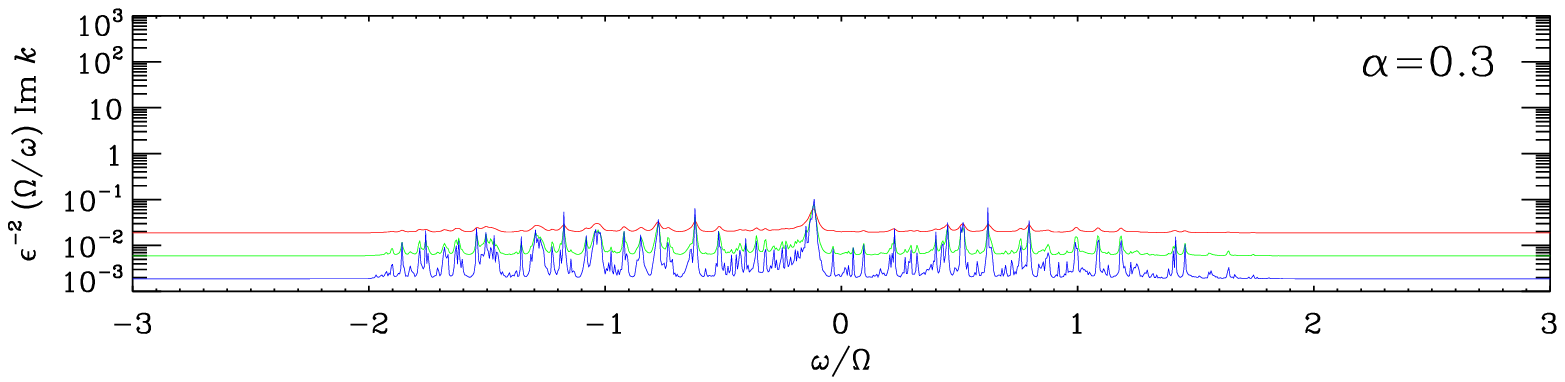}}
\centerline{\epsfbox{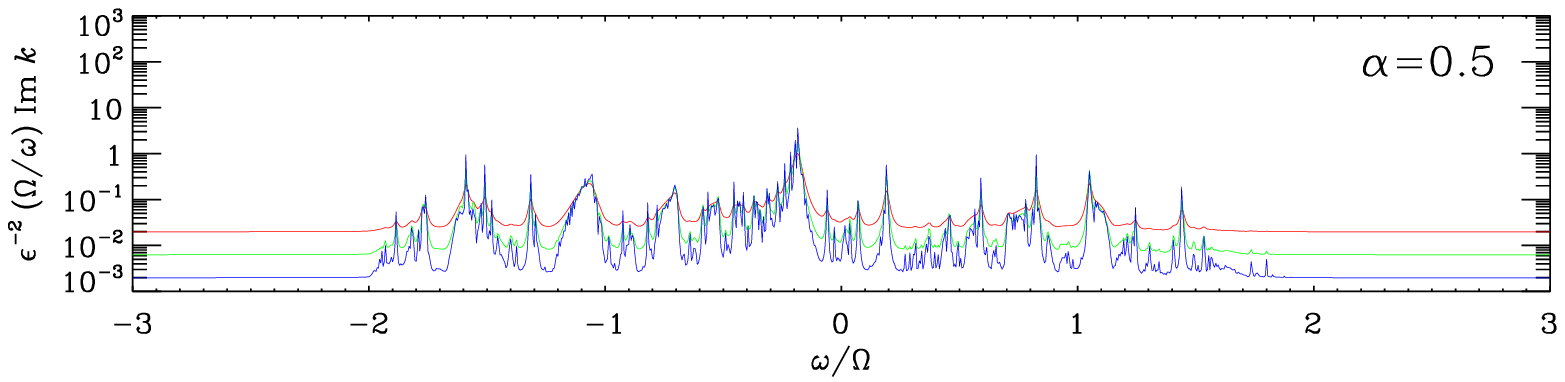}}
\centerline{\epsfbox{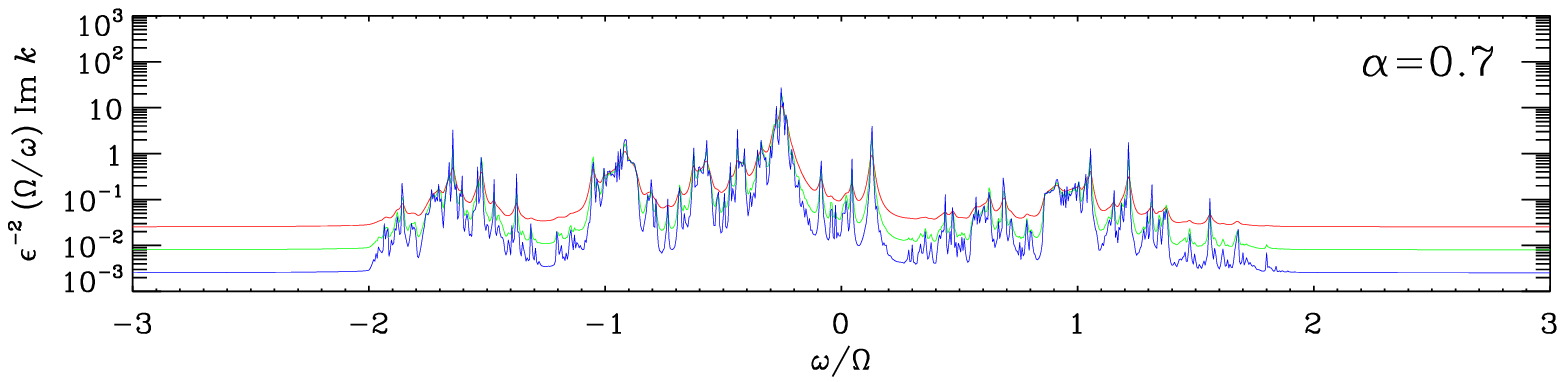}}
\centerline{\epsfbox{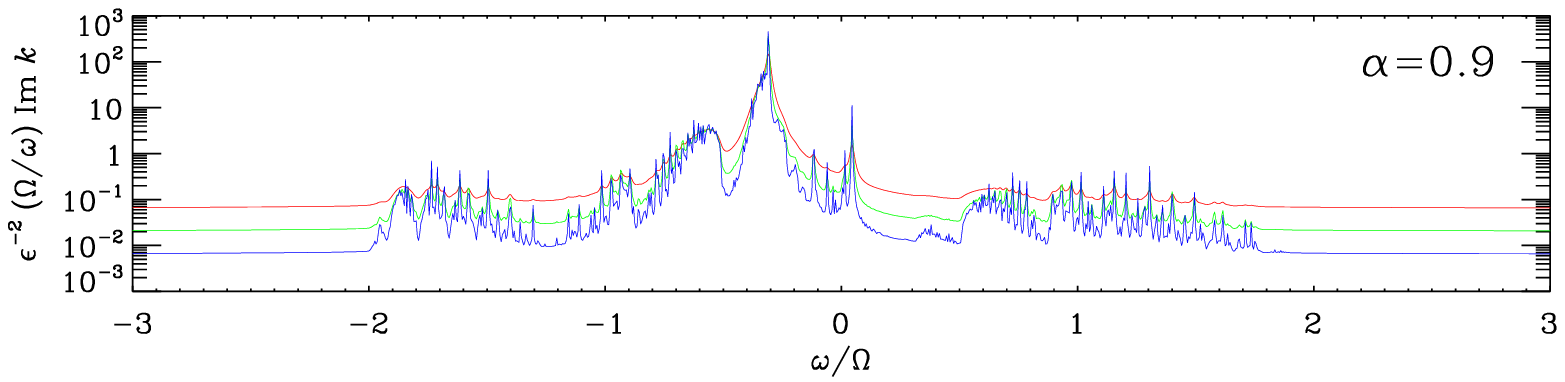}}
\caption{Frequency-dependent tidal response to $Y_2^2$ of a
  homogeneous body with an incompressible fluid envelope and a
  perfectly rigid solid core of fractional radius $\alpha=0.1$ (top
  panel), $0.3$, $0.5$, $0.7$ or $0.9$ (bottom panel).  As in
  \citet{2009MNRAS.396..794O} the fluid is inviscid but experiences a
  frictional force with a damping coefficient $\gamma$ given by
  $\gamma/\Omega=10^{-2}$ (red line), $10^{-2.5}$ (green line) or
  $10^{-3}$ (blue line); also, the surface is not free but has a
  prescribed radial motion.  The imaginary part of the Love number is
  deduced from the dissipation rate and is plotted on a logarithmic
  scale after multiplication by $\epsilon^{-2}(\Omega/\omega)$, where
  $\Omega=\epsilon(GM/R^3)^{1/2}$.  A numerical resolution of
  $L=N=400$ is used.}
\label{f:imk_l2_m2_homogeneous}
\end{figure*}

\begin{figure}
\centerline{\epsfbox{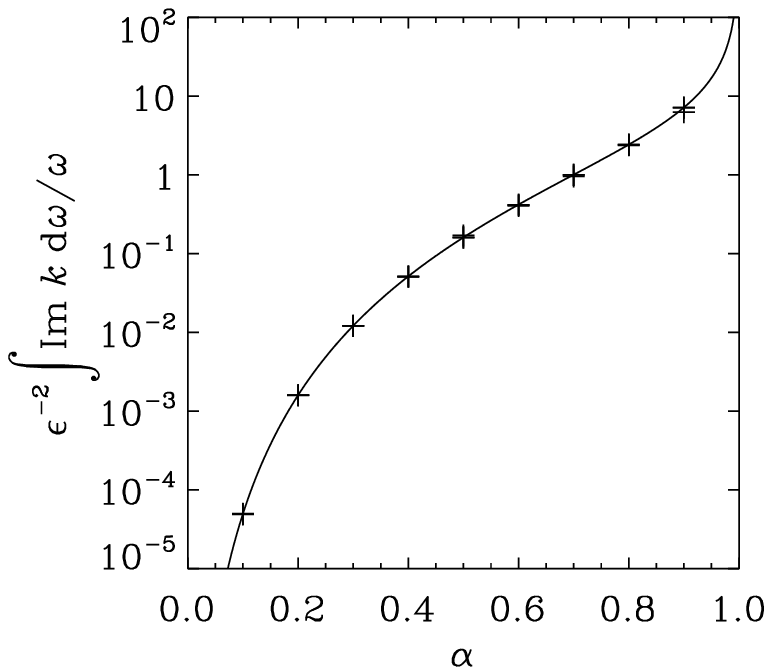}}
\caption{Frequency-integral of the imaginary part of the Love number,
  weighted by $1/\omega$ and divided by $\epsilon^2$, for a
  homogeneous body, versus fractional core size.  The solid line shows
  the analytical result~(\ref{int_l2_m2_homogeneous}) of the impulse
  calculation, while the symbols show the numerical integrals based on
  Figure~\ref{f:imk_l2_m2_homogeneous} and similar calculations.}
\label{f:int_l2_m2_homogeneous}
\end{figure}

For $l=m=2$ we have
\begin{eqnarray}
  \hat E&=&\f{80}{189}\alpha^5(1-\alpha^5)|C|^2\rho R\Omega^2\nonumber\\\
  &=&\f{20}{189}\alpha^5(1-\alpha^5)^{-1}|A+B|^2\f{\rho R^3\Omega^2}{g^2},
\end{eqnarray}
which is proportional to $\alpha^5$ for small $\alpha$.  This strong
dependence on core size agrees with the behaviour of the
(frequency-averaged) dissipation rate found by
\citet{2009ApJ...696.2054G}, \citet{2009MNRAS.396..794O} and
\citet{2010JFM...643..363R}, although \citet{2009ApJ...696.2054G}
considered a nonlinear mechanism involving wave breaking.  In the
simplest case $\bar\rho=\rho$, we have $(A+B)/A=(2l+1)/(2l-2)=5/2$ and
so
\begin{equation}
  \int_{-\infty}^\infty\mathrm{Im}[K_2^2(\omega)]\,\f{\rmd\omega}{\omega}=\f{100\pi}{63}\epsilon^2\left(\f{\alpha^5}{1-\alpha^5}\right).
\label{int_l2_m2_homogeneous}
\end{equation}
We have verified this result numerically as follows.  In
Figure~\ref{f:imk_l2_m2_homogeneous} we show the frequency-dependent
response to $l=m=2$ tidal forcing for a homogeneous body with an
incompressible fluid envelope and a perfectly rigid solid core of
various sizes.  The tidal response is computed as in
\citet{2009MNRAS.396..794O} and this figure is equivalent to (parts
of) Figures~1 and~2 of the earlier paper except that the energy
dissipation rate is converted into the imaginary part of the Love
number.  This conversion brings in the parameter $\epsilon$, defined
in equation~(\ref{epsilon}), which is unspecified in these
calculations but is assumed to be small; as noted above, we expect
$\mathrm{Im}\,k\propto\epsilon^2$ for inertial waves.  Then the
integral in equation~(\ref{int_l2_m2_homogeneous}) is computed
numerically after subtracting the baseline dissipation rate for each
curve (which corresponds to the frictional damping of the non-wavelike
part of the tide).  The results are plotted in
Figure~\ref{f:int_l2_m2_homogeneous} for nine values of $\alpha$ and
three values of the frictional damping coefficient.  Very good
agreement is found with equation~(\ref{int_l2_m2_homogeneous}).

The divergence that occurs as $\alpha\to1$ deserves some comment.  The
coefficient $C$ diverges in this limit, as do the non-radial
components of the non-wavelike displacement.  The thin, incompressible
fluid shell is squeezed radially by the tidal force and produces rapid
horizontal motion.  Through the Coriolis force, a large wavelike
velocity is also generated.  This behaviour would be weakened if the
core were not perfectly rigid but instead underwent some tidal
deformation.

\begin{figure*}
\centerline{\epsfbox{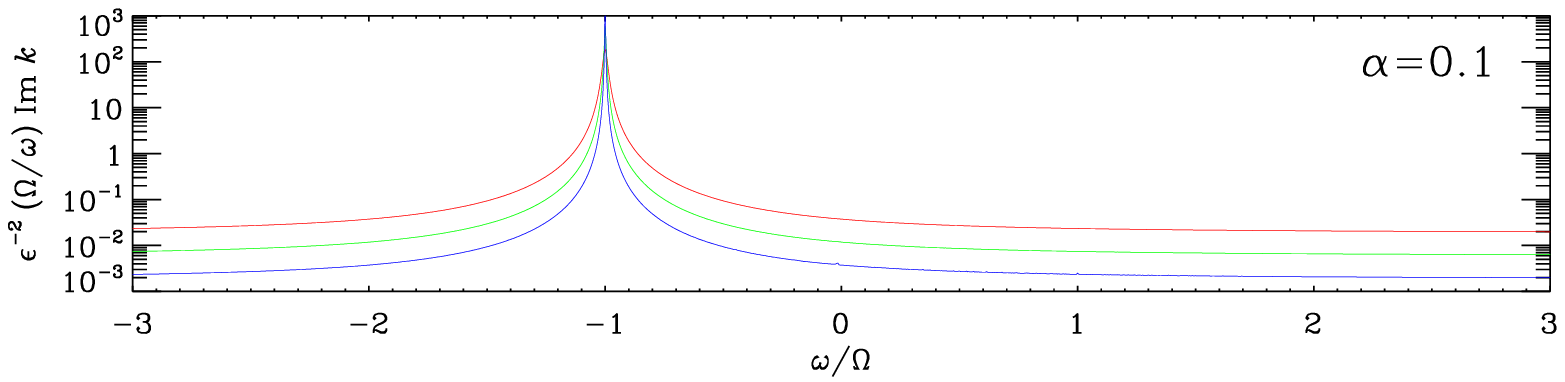}}
\centerline{\epsfbox{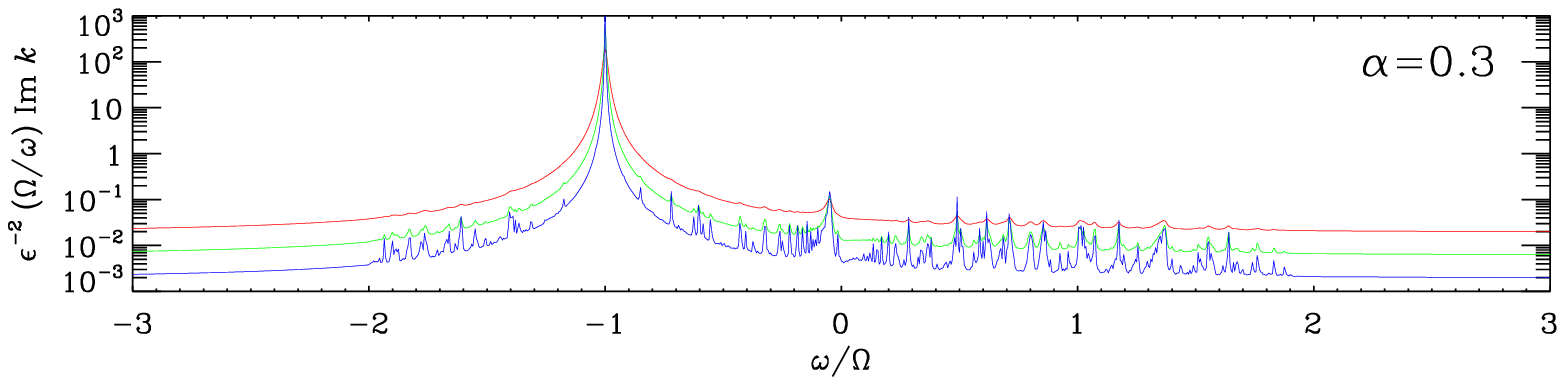}}
\centerline{\epsfbox{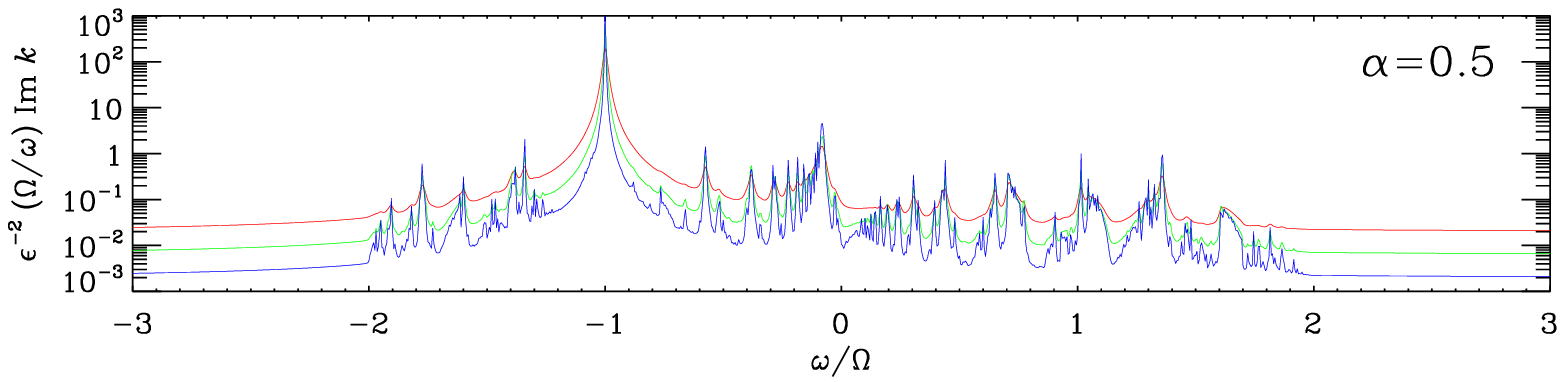}}
\centerline{\epsfbox{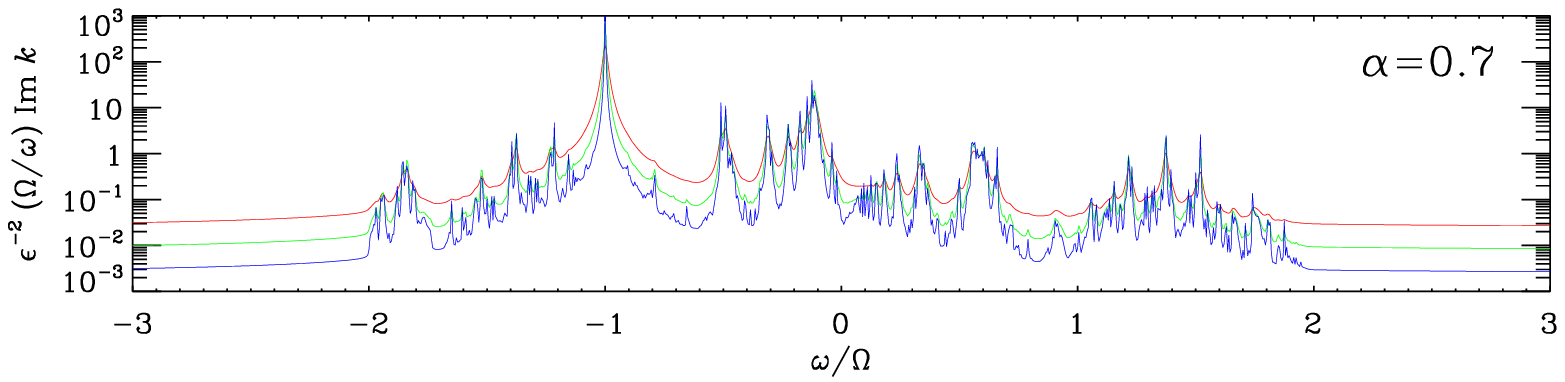}}
\centerline{\epsfbox{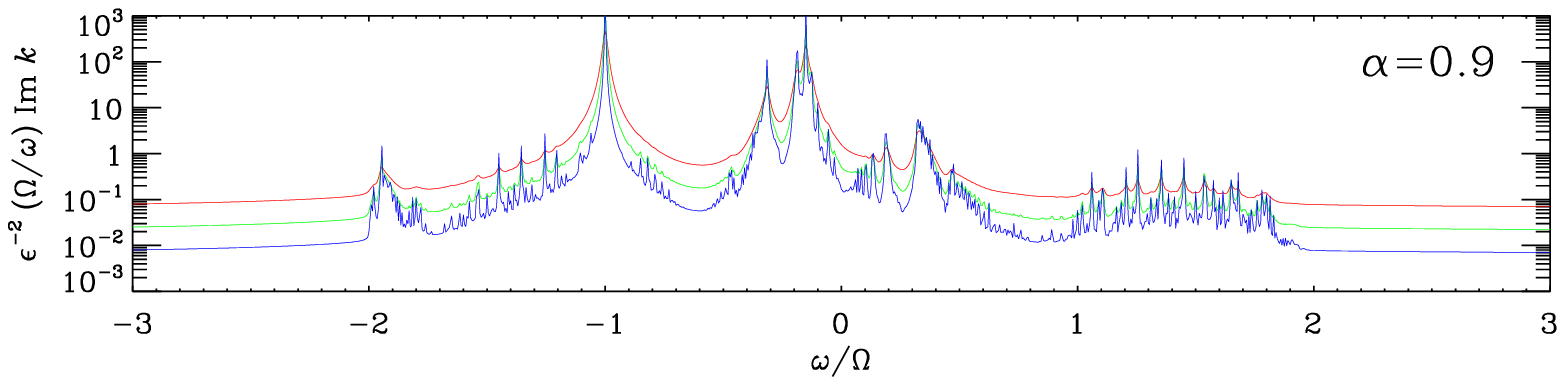}}
\caption{As for Figure~\ref{f:imk_l2_m2_homogeneous} but for the tidal potential $Y_2^1$.  There is a formal resonance with the spin-over mode at $\omega/\Omega=-1$.}
\label{f:imk_l2_m1_homogeneous}
\end{figure*}

\begin{figure*}
\centerline{\epsfbox{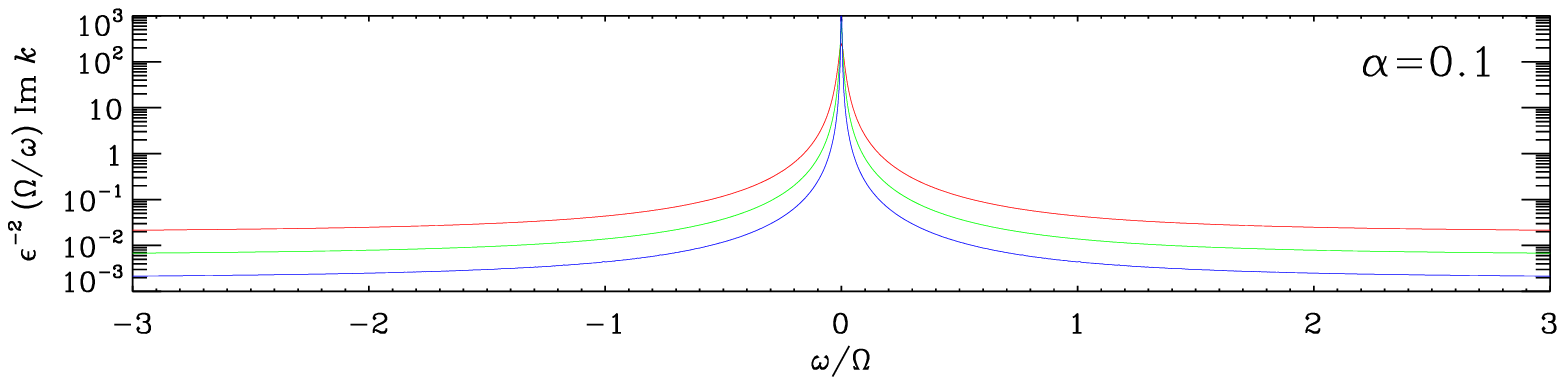}}
\centerline{\epsfbox{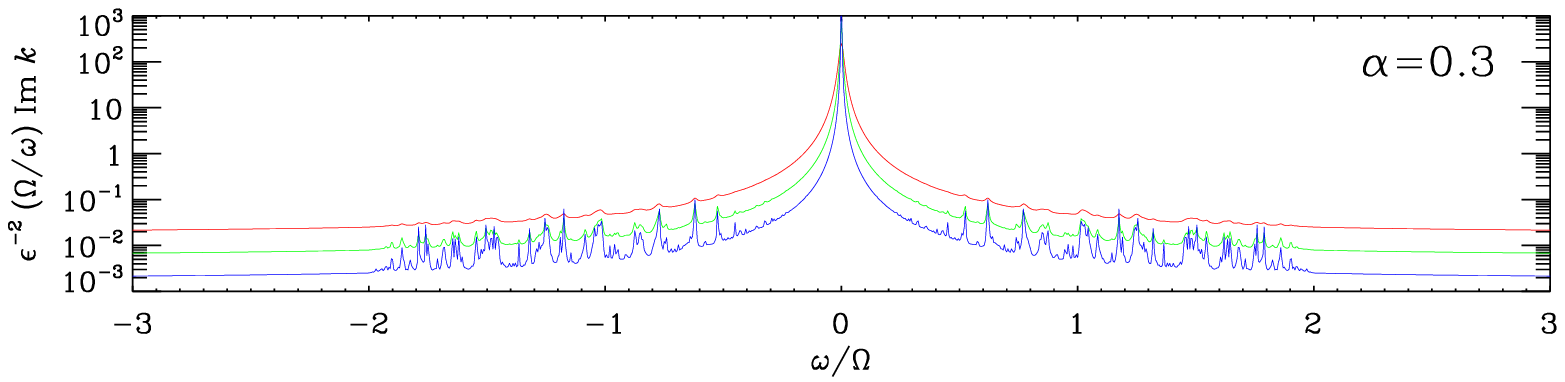}}
\centerline{\epsfbox{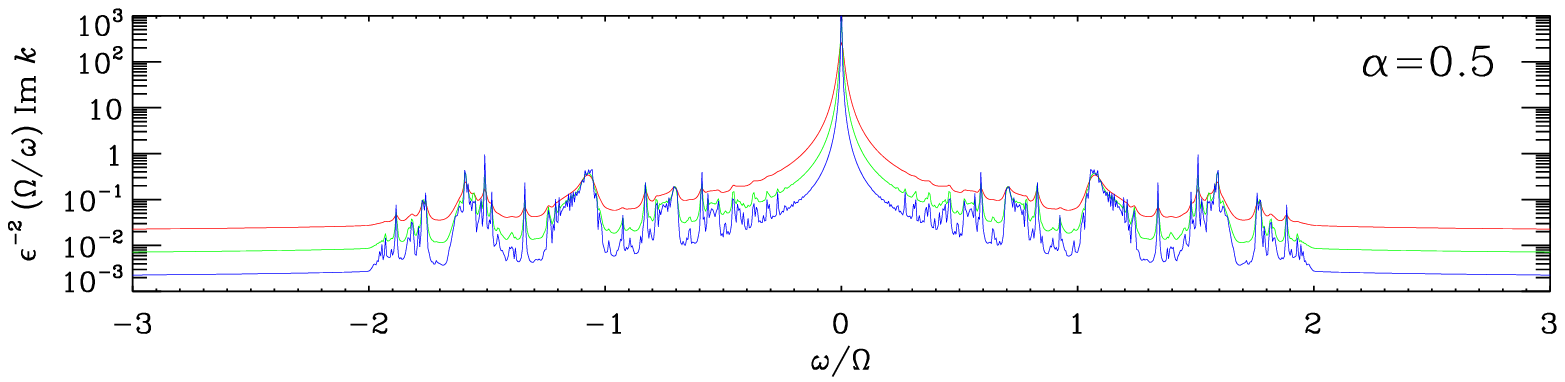}}
\centerline{\epsfbox{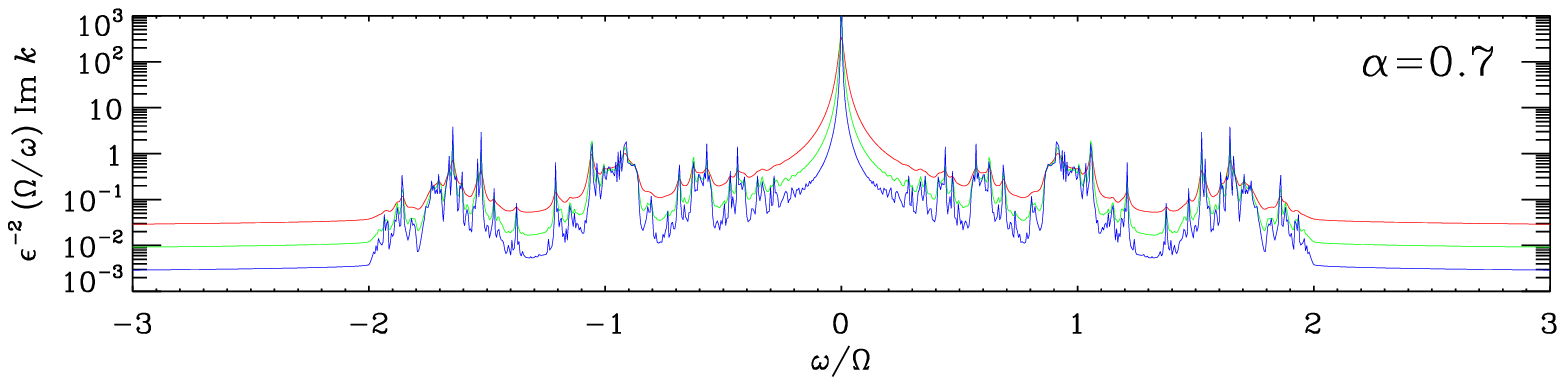}}
\centerline{\epsfbox{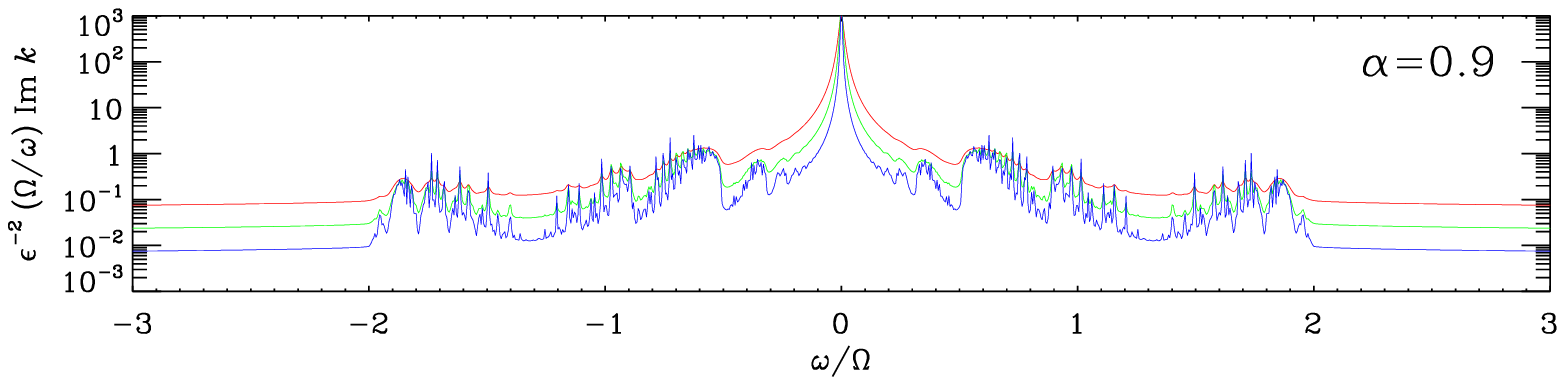}}
\caption{As for Figure~\ref{f:imk_l2_m2_homogeneous} but for the tidal potential $Y_2^0$.  There is a formal resonance with the spin-up mode at $\omega/\Omega=0$.}
\label{f:imk_l2_m0_homogeneous}
\end{figure*}

Generalizations of equation~(\ref{int_l2_m2_homogeneous}) can be found
in Appendix~\ref{a:analytical}.  In particular,
\begin{equation}
  \int_{-\infty}^\infty\mathrm{Im}[K_2^1(\omega)]\,\f{\rmd\omega}{\omega}=\f{5\pi}{504}\epsilon^2\left(\f{189+256\alpha^5}{1-\alpha^5}\right)
\end{equation}
and
\begin{equation}
  \int_{-\infty}^\infty\mathrm{Im}[K_2^0(\omega)]\,\f{\rmd\omega}{\omega}=\f{5\pi}{14}\epsilon^2\left(\f{7+8\alpha^5}{1-\alpha^5}\right).
\end{equation}
Note that, although these expressions are increasing functions of
$\alpha$, they do not vanish at $\alpha=0$.  We have also found
excellent agreement between these expressions and the numerically
integrated Love numbers using the method described above for the case
$l=m=2$.  The frequency-dependent responses are illustrated in
Figures~\ref{f:imk_l2_m1_homogeneous}
and~\ref{f:imk_l2_m0_homogeneous}.  Note that the enhancement of the
tidal response in these cases results from resonances with the
spin-over mode ($m=1$, $\omega/\Omega=-1$) and the spin-up mode
($m=0$, $\omega/\Omega=0$).  These are trivial modes and there should
be no associated dissipation.  In these calculations the dissipation
is artificial and results from the use of a frictional, rather than
viscous, force.  In either case, however, there is a formal
singularity in $\mathrm{Im}\,k$ that gives a non-vanishing
contribution to the integral even when $\alpha=0$.  In a viscous fluid
the resonance should be of zero width.

\begin{figure*}
\centerline{\epsfbox{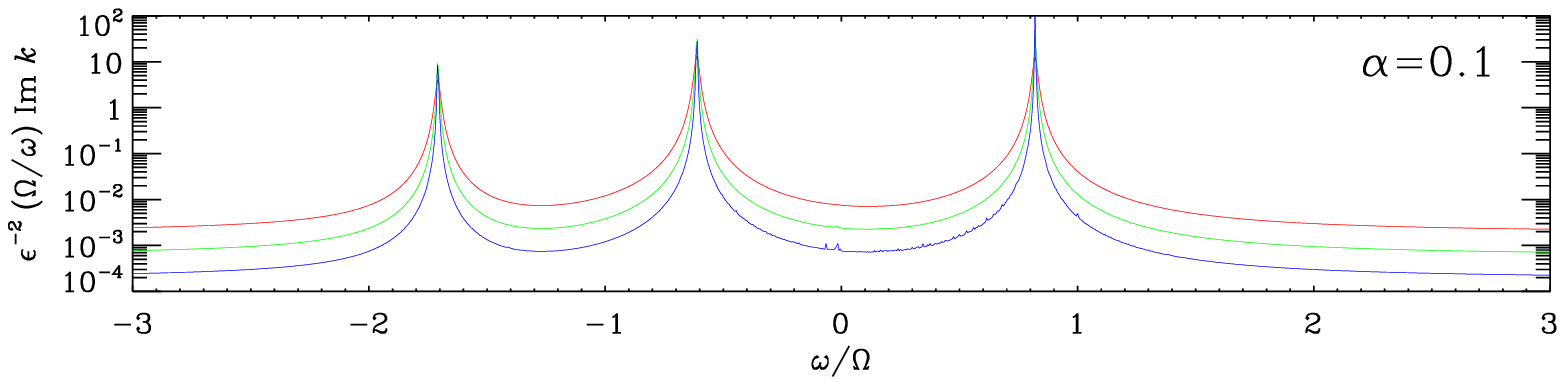}}
\centerline{\epsfbox{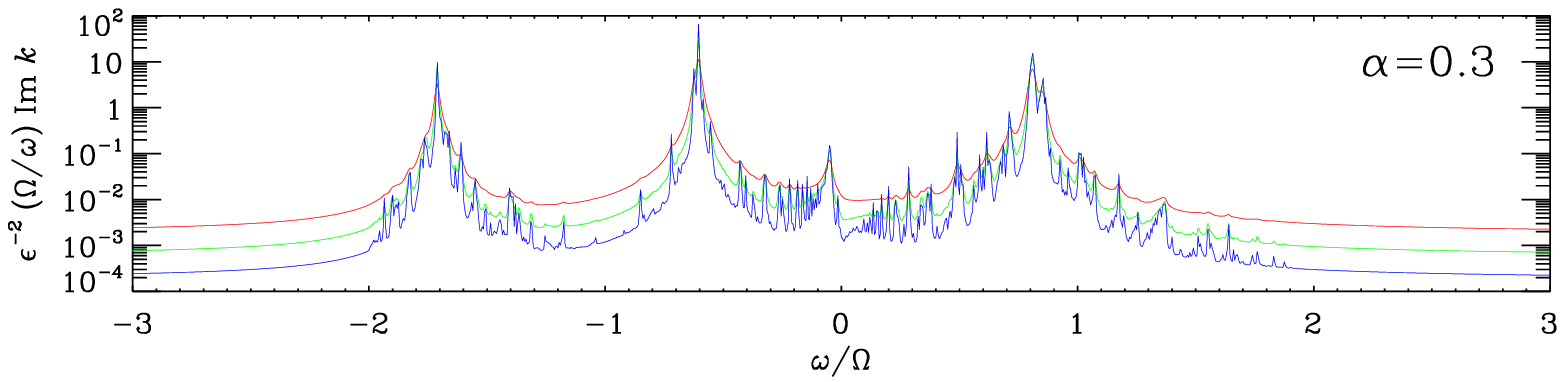}}
\centerline{\epsfbox{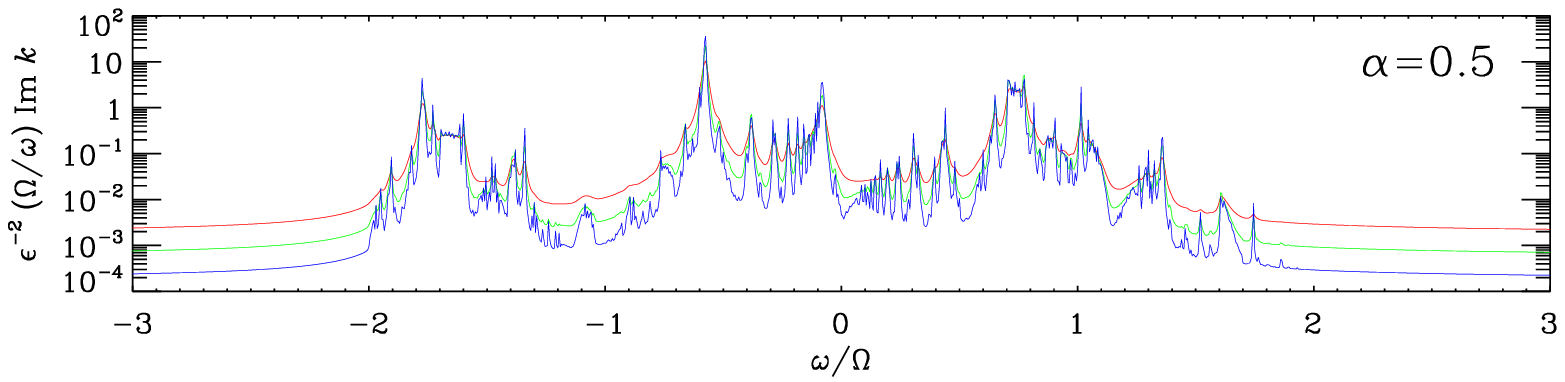}}
\centerline{\epsfbox{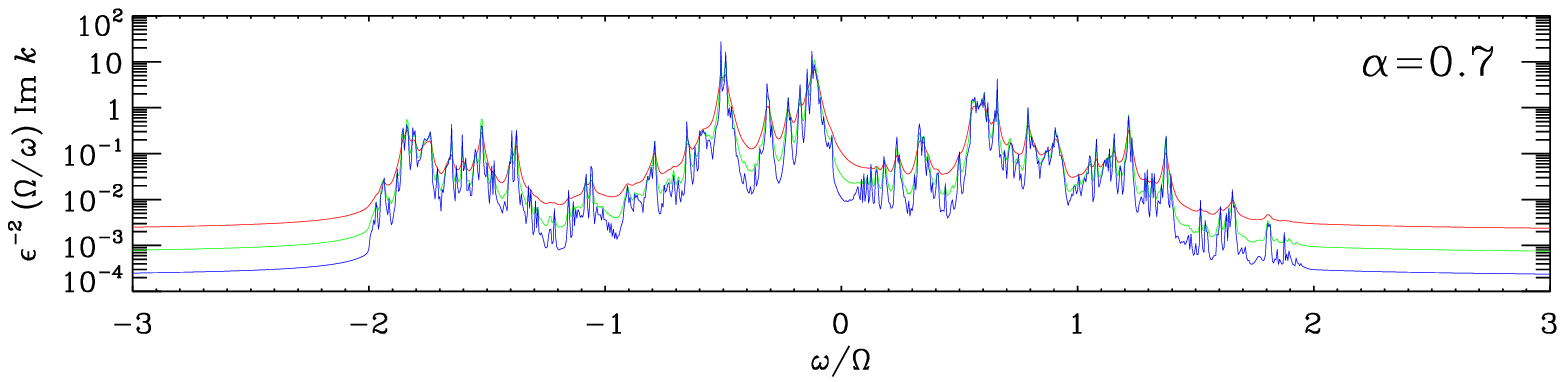}}
\centerline{\epsfbox{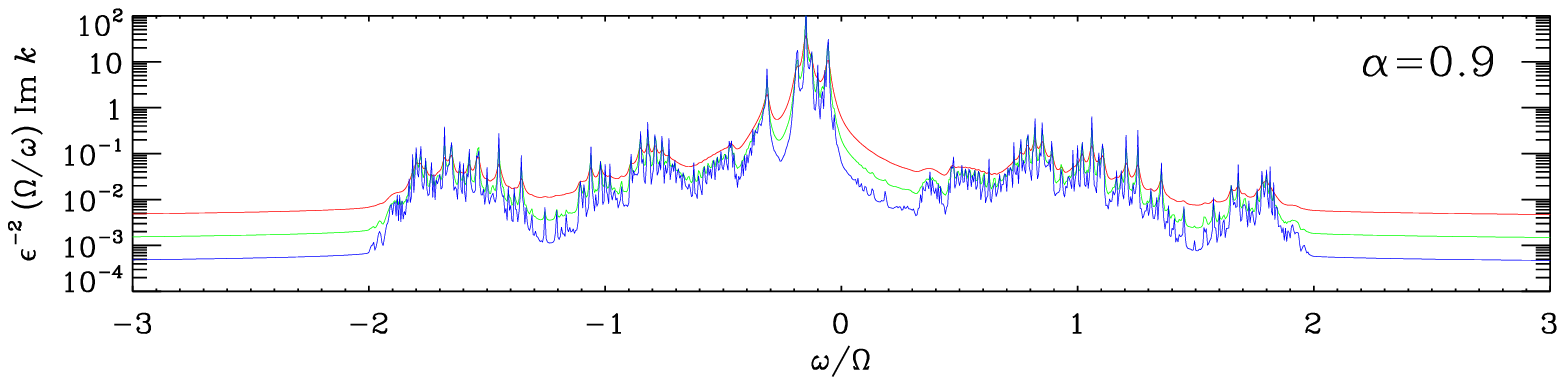}}
\caption{As for Figure~\ref{f:imk_l2_m2_homogeneous} but for the tidal potential $Y_4^1$.  This case is chosen to illustrate the richer response (for small core sizes) to tesseral harmonics of higher degrees.  Note, however, the different scale on the vertical axis.}
\label{f:imk_l4_m1_homogeneous}
\end{figure*}

To illustrate the richer possibilities that exist for tidal potential
components beyond the quadrupolar ones, we show similar results for
$l=4$ and $m=1$ in Figure~\ref{f:imk_l4_m1_homogeneous}.  In this case
the response for small values of $\alpha$ is dominated by resonances
with non-trivial inertial modes of the full sphere (at
$\omega/\Omega=-1.7080$, $-0.6120$ and $+0.8200$, as expected from
equation~\ref{cubic}).  For larger core sizes the frequency-dependence
is qualitatively similar to that seen for quadrupolar tidal
potentials.  The analytical result in this case is
\begin{equation}
  \int_{-\infty}^\infty\mathrm{Im}[K_4^1(\omega)]\,\f{\rmd\omega}{\omega}=\f{27\pi}{616000}\epsilon^2\left(\f{6875+7168\alpha^9}{1-\alpha^9}\right).
\end{equation}

\begin{figure}
\centerline{\epsfbox{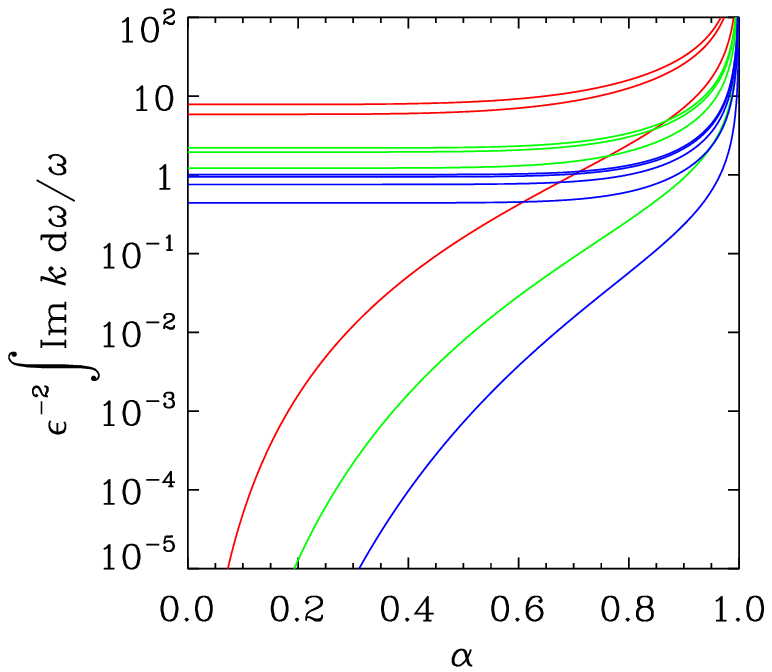}}
\caption{As for Figure~\ref{f:int_l2_m2_homogeneous} but showing only the results of the impulse calculation for a homogeneous body and for various tidal potentials with $l=2$ (red lines), $l=3$ (green lines) and $l=4$ (blue lines).  On the left of the figure the ordering of curves is, from top to bottom: $Y_2^0$,
$Y_2^1$, $Y_3^0$, $Y_3^1$, $Y_3^2$, $Y_4^0$, $Y_4^1$, $Y_4^2$, $Y_4^3$, $Y_2^2$, $Y_3^3$, $Y_4^4$.}
\label{f:int_various}
\end{figure}

The integrated responses for all tidal potentials up to $l=4$ are shown in Figure~\ref{f:int_various}.  Although there is a general weakening of the response with increasing $l$, the integrated response to a tesseral harmonic of higher $l$ is usually much stronger than that to a sectoral harmonic of lower $l$.

\subsection{Application to a piecewise-homogeneous fluid model}
\label{s:fluid core}

We now examine the effect of replacing the rigid solid core with a
fluid core.  We therefore consider a homogeneous fluid of density
$\rho_1$ in $0<r<\alpha R$ and another of density $\rho_2$ in $\alpha
R<r<R$, with $\rho_1\ge\rho_2$ for gravitational stability.

At an interface, where the density changes abruptly,
$\xi_{\rmnw,r}=-(\Phi'+\Psi)/g$.  This is because the Lagrangian
pressure perturbation $p'-\rho g\xi_r$ is continuous, but
$p'=\rho(W-\Phi'-\Psi)$ and $W$ is supposed to be negligible
[$O(\epsilon^2)$] in the low-frequency approximation.  If $\rho$ is
discontinuous then we find the stated condition, which means that the
interface moves equipotentially.  (This should be valid provided that
the tidal frequency is small compared to the relevant interfacial
gravity wave frequency.)

The Helmholtz-like equation is then
\begin{equation}
  \nabla^2\hat\Phi'+\f{4\pi G}{g}(\hat\Phi'+\hat\Psi)[(\rho_1-\rho_2)\delta(r-\alpha R)+\rho_2\delta(r-R)]=0.
\end{equation}
Given $\hat\Psi=A(r/R)^lY_l^m$, the solution is $\hat\Phi'=\hat\Phi'_l(r)Y_l^m$, with
\begin{equation}
  \hat\Phi'_l=\left\{
  \begin{array}{ll}
    (B_1+B_2)(r/R)^l,&0<r<\alpha R,\\
    B_1(r/R)^l+B_2\alpha^{2l+1}(R/r)^{l+1},&\alpha R<r<R,\\
    (B_1+B_2\alpha^{2l+1})(R/r)^{l+1},&r>R,
  \end{array}
  \right.
\end{equation}
where the coefficients satisfy the matching conditions
\begin{equation}
  (2l+1)B_2=3(1-f)(A+B_1+B_2),
\end{equation}
\begin{equation}
  (2l+1)B_1=\f{3f}{f+(1-f)\alpha^3}(A+B_1+B_2\alpha^{2l+1}),
\end{equation}  
with $f=\rho_2/\rho_1$.  We shall not write explicitly the algebraic
solution for $B_1$ and $B_2$.

$\hat X$ satisfies Laplace's equation in each fluid shell and the
solution (regular at $r=0$) is of the form
\begin{equation}
  \hat X_l=\left\{
  \begin{array}{ll}
    C_1(r/R)^l,&0<r<\alpha R,\\
    C_2(r/R)^l+C_3\alpha^{2l+1}(R/r)^{l+1},&\alpha R<r<R.
  \end{array}
  \right.
\end{equation}
The condition on the radial displacement is $\rmd\hat X_l/\rmd
r=(\hat\Phi_l+\hat\Psi_l)/g$ at each interface.  Thus
\begin{equation}
  lC_1=lC_2-(l+1)C_3=\f{3(A+B_1+B_2)}{4\pi G\rho_1},
\end{equation}
\begin{equation}
  lC_2-(l+1)C_3\alpha^{2l+1}=\f{3(A+B_1+B_2\alpha^{2l+1})}{4\pi G(\rho_2+(\rho_1-\rho_2)\alpha^3)}.
\end{equation}
The fact that $C_3\ne0$ in general means that $\hat X_l$ is
discontinuous in general at $r=\alpha R$.  This means that the
non-wavelike displacement has a tangential discontinuity.  The
singularity of vorticity results from the non-barotropic nature of the
density jump.

$\hat W$ also satisfies Laplace's equation and the solution is
\begin{equation}
  \hat W_l=\left\{
  \begin{array}{ll}
    -2\rmi m\Omega(C_1/l)(r/R)^l,\\
    -2\rmi m\Omega[(C_2/l)(r/R)^l-(C_3/(l+1))\alpha^{2l+1}(R/r)^{l+1}],
  \end{array}
  \right.
\end{equation}
such that
\begin{equation}
  \hat a_l=\hat b_l=0.
\end{equation}
(Here and below, the first and second cases of the expressions in braces apply for $0<r<\alpha R$ and $\alpha R<r<R$, respectively.)
The toroidal part is given by
\begin{equation}
  \hat c_{l-1}=\left\{
  \begin{array}{ll}
    -2\Omega\tilde q_l r^{-2}(2l+1)C_1(r/R)^l,\\
    -2\Omega\tilde q_l r^{-2}(2l+1)C_2(r/R)^l,
  \end{array}
  \right.
\end{equation}
\begin{equation}
  \hat c_{l+1}=\left\{
  \begin{array}{ll}
    0,\\
    -2\Omega\tilde q_{l+1}r^{-2}(2l+1)C_3\alpha^{2l+1}(R/r)^{l+1}.
  \end{array}
  \right.
\end{equation}
The contributions to the impulse energy are then
\begin{equation}
  \hat E_l=0,
\end{equation}
\begin{eqnarray}
  \hat E_{l-1}&=&l(l-1)(2l+1)\Omega^2\tilde q_l^2\nonumber\\
  &&\times[\rho_1\alpha^{2l+1}|C_1|^2+\rho_2(1-\alpha^{2l+1})|C_2|^2]R,
\end{eqnarray}
\begin{eqnarray}
  \hat E_{l+1}&=&(l+1)(l+2)(2l+1)\Omega^2\tilde q_{l+1}^2\nonumber\\
  &&\times\rho_2\alpha^{2l+1}(1-\alpha^{2l+1})|C_3|^2R.
\end{eqnarray}

\begin{figure}
\centerline{\epsfbox{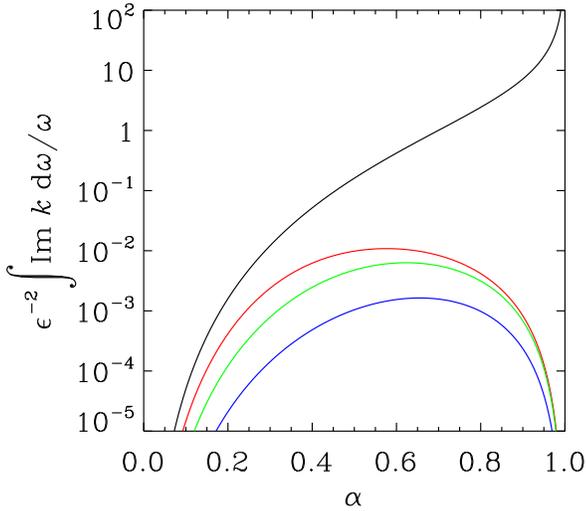}}
\caption{As for Figure~\ref{f:int_l2_m2_homogeneous} but showing only the results of the impulse calculation for a homogeneous body (black line) and for the piecewise-homogeneous fluid model with density ratio $f=0.25$ (red line), $f=0.5$ (green line) or $f=0.75$ (blue line).}
\label{f:int_l2_m2_twofluid}
\end{figure}

These results are converted into integrals of $\mathrm{Im}\,k$ using
equation~(\ref{deltae}) and plotted in
Figure~\ref{f:int_l2_m2_twofluid} for various values of~$f$ in the
case $l=m=2$; the analytical result is given in equation~(\ref{pwh}).
For small values of~$f$ (large density contrasts) the impulsive
response is similar to, but weaker than, the case of a homogeneous
fluid with a solid core; the differences are greater when the core is
larger.  These results show that a fluid core can play a similar (if
weaker) role in the excitation of inertial waves to that of a rigid
solid core.  As $f$ tends to~$1$ the response disappears because the
system becomes a full homogeneous sphere.

\subsection{Application to polytropes}

\begin{figure*}
\centerline{\epsfbox{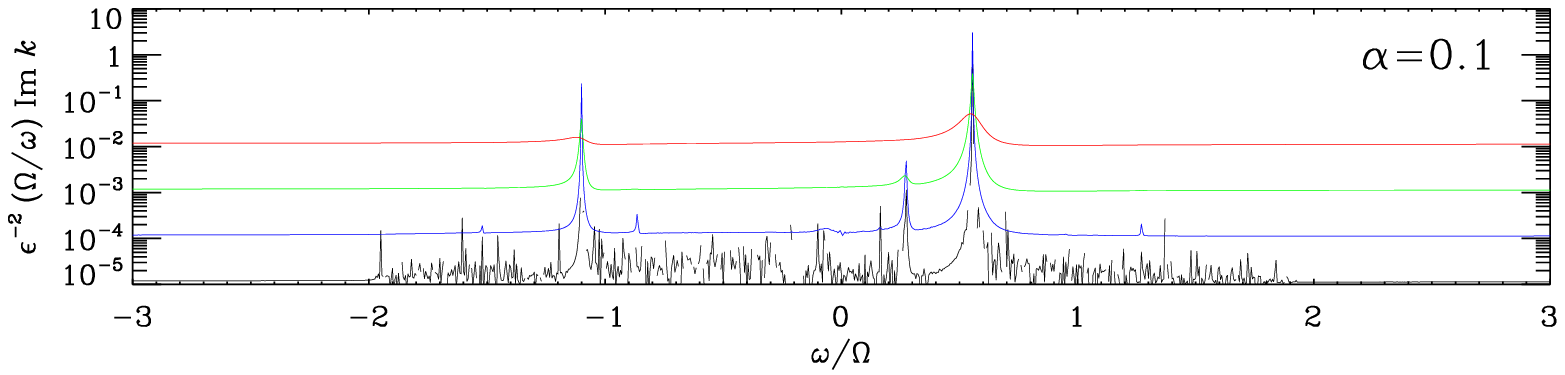}}
\centerline{\epsfbox{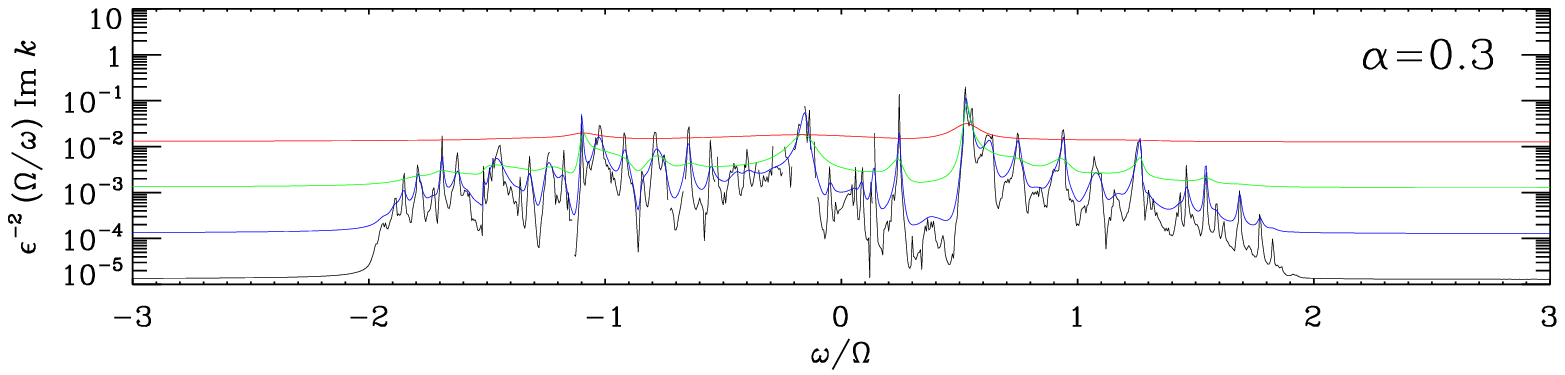}}
\centerline{\epsfbox{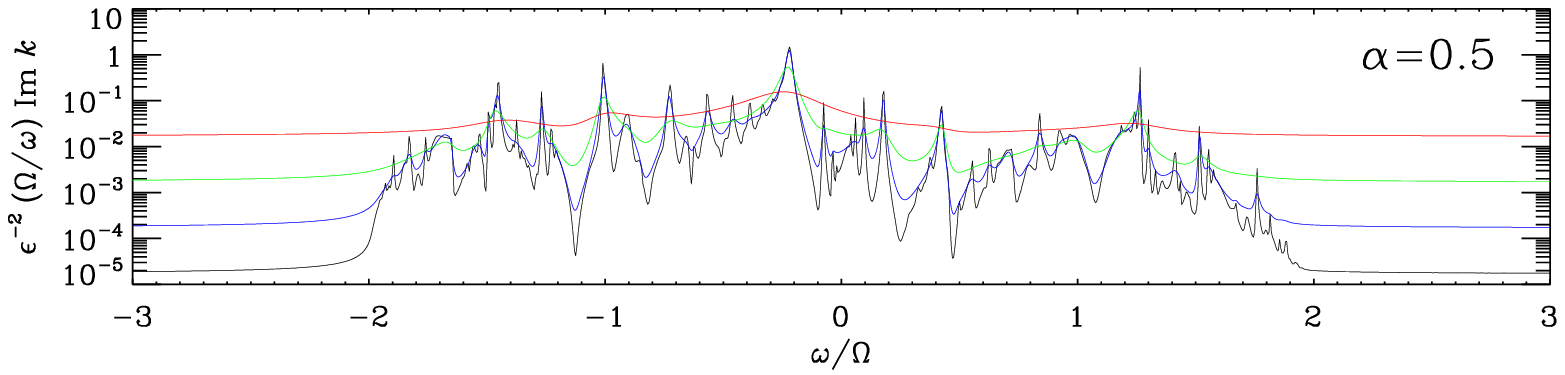}}
\centerline{\epsfbox{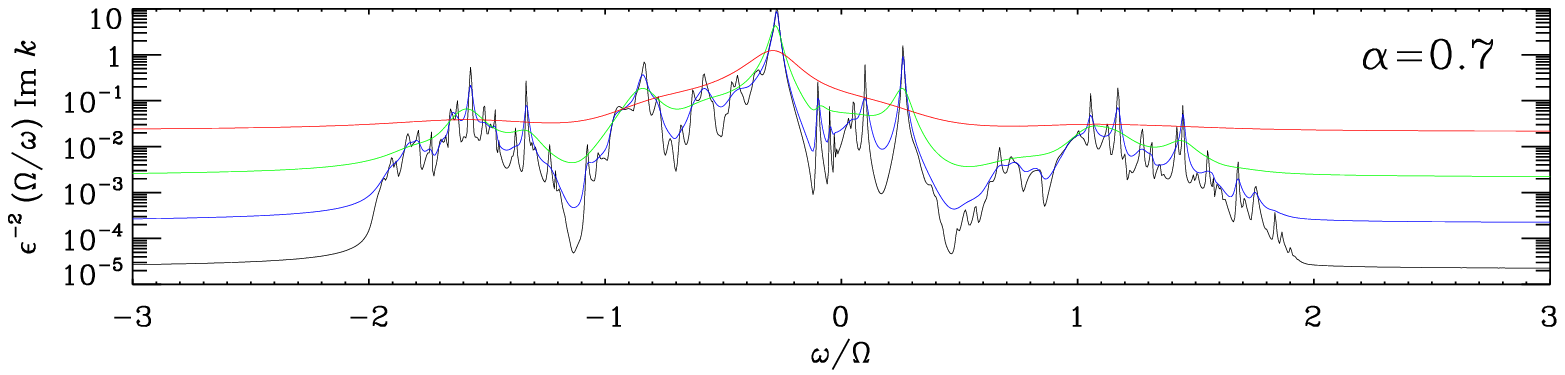}}
\centerline{\epsfbox{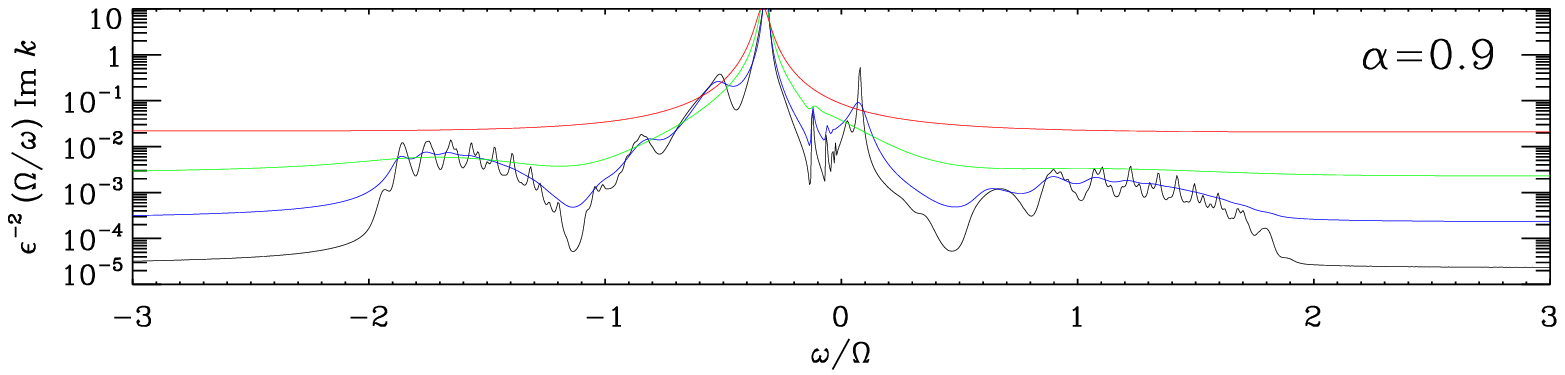}}
\caption{Frequency-dependent tidal response to $Y_2^2$ of an $n=1$
  polytrope with a fluid envelope and a perfectly rigid solid core of
  fractional radius $\alpha=0.1$ (top panel), $0.3$, $0.5$, $0.7$ or
  $0.9$ (bottom panel).  The fluid is viscous and the Ekman number is
  $\mathrm{Ek}=10^{-3}$ (red line), $10^{-4}$ (green line), $10^{-5}$
  (blue line) or $10^{-6}$ (black line).  The imaginary part of the
  Love number is computed directly and is plotted on a logarithmic
  scale after multiplication by $\epsilon^{-2}(\Omega/\omega)$, where
  $\Omega=\epsilon(GM/R^3)^{1/2}$.  In these calculations
  $\epsilon=0.01$ and the fluid envelope is truncated with a free
  surface at a fractional radius of $0.99$.  A numerical resolution of
  $L=N=200$ is used.  The results for $\mathrm{Ek}=10^{-6}$ are not
  fully converged at this resolution and occasionally return negative
  values of $\epsilon^{-2}(\Omega/\omega)\mathrm{Im}\,k$, resulting in
  gaps in the black line.}
\label{f:imk_l2_m2_polytrope}
\end{figure*}

We have also computed the impulsive energy transfer $\hat E$ for
polytropes of various indices $n$.  This is done by first solving the
Lane--Emden equation for the structure of the polytrope, then
computing the impulsive response by solving the ODEs of
Section~\ref{s:impulse} numerically by standard methods.  We allow for
a perfectly rigid solid core to be present, and assume that the mass
of the core is equal to the mass of fluid that would occupy the
equivalent region in a full polytrope.

The frequency-dependent response of an $n=1$ polytrope to $l=m=2$
tidal forcing is illustrated in Figure~\ref{f:imk_l2_m2_polytrope}.
These results are similar to those presented by
\citet{2004ApJ...610..477O} but are computed by a different numerical
method which will be described in forthcoming work (Ogilvie, in
preparation).  Rather than making a low-frequency approximation as in
\citet{2004ApJ...610..477O}, we solve the full set of linearized
equations for a compressible, self-gravitating and viscous fluid while
neglecting centrifugal effects (the parameter $\epsilon$ is here set
to $0.01$).  This method has the advantage that the Love number can be
measured directly, because the gravitational potential perturbation is
part of the numerical solution.  The fluid is subject to viscous,
rather than frictional, forces, as quantified by the Ekman number
$\mathrm{Ek}=\nu/(2\Omega R^2)$, where $\nu$ is the kinematic
viscosity.  In all cases except those at the lowest Ekman number
$\mathrm{Ek}=10^{-6}$, where the numerical resolution may be
insufficient, the imaginary part of the Love number is consistent with
the viscous dissipation rate to a high degree of accuracy.

\begin{figure}
\centerline{\epsfbox{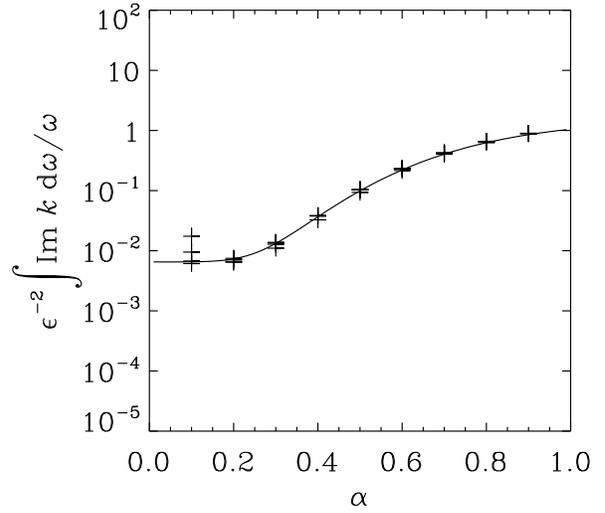}}
\caption{As for Figure~\ref{f:int_l2_m2_homogeneous} but for an $n=1$ polytrope.  The numerical integrals are based on Figure~\ref{f:imk_l2_m2_polytrope} and similar calculations.}
\label{f:int_l2_m2_polytrope}
\end{figure}

\begin{figure}
\centerline{\epsfbox{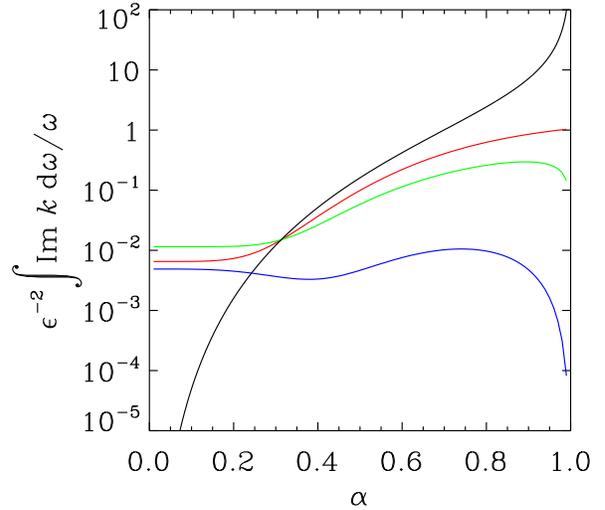}}
\caption{As for Figure~\ref{f:int_l2_m2_homogeneous} but showing only the results of the impulse calculation for a homogeneous body (black line) and for polytropes of indices $n=1$ (red line), $n=3/2$ (green line) and $n=3$ polytrope (blue line).}
\label{f:int_l2_m2_various}
\end{figure}

While there are many qualitative features in common between
Figures~\ref{f:imk_l2_m2_homogeneous} and~\ref{f:imk_l2_m2_polytrope},
there are also important differences.  For very small core sizes such
as $\alpha=0.1$, the homogeneous body exhibits almost no response to
$Y_2^2$, while the $n=1$ polytrope shows a few resonant peaks.  These
correspond to the resonant excitation of inertial normal modes of the
full polytrope, which have been computed by
\citet{1999ApJ...521..764L}.\footnote{Visible in the top panel of
  Figure~\ref{f:imk_l2_m2_polytrope}, for example, are the frequencies
  $\omega/\Omega=0.5566$ and $-1.1000$ listed in the fifth column of
  their Table~6, albeit with a different sign convention.}  Similar
modes also exist in the homogeneous sphere, but they have no overlap
with the $Y_2^2$ tidal potential and are not excited.  For larger core
sizes, normal modes cease to be important, until the limit of a thin
shell is reached, in which the dominant feature is the excitation of a
planetary wave/Rossby wave/r~mode at $\omega/\Omega=-1/3$
\citep{2009MNRAS.396..794O}.

In Figure~\ref{f:int_l2_m2_polytrope} we show the integrated quantity
$\int\mathrm{Im}\,k\,\rmd\omega/\omega$, divided by $\epsilon^2$, as
determined both from numerical integration of the curves in
Figure~\ref{f:imk_l2_m2_polytrope} (and similar calculations) and from
the impulse calculation described above.  The two are in very good
agreement, except in a few cases where the numerical integration
probably has insufficient frequency-resolution to capture the narrow
peaks that occur in $\mathrm{Im}\,k$ at low $\mathrm{Ek}$, or where
the numerical resolution is insufficient to determine the dissipation
rate accurately.

Figure~\ref{f:int_l2_m2_various} compares similar results (but without
the numerically integrated values) for a homogeneous body and for
polytropes of indices $n=1$, $3/2$ and $3$.  This figure shows that,
as the polytropic index increases and the body becomes more centrally
condensed, the integrated tidal response to $Y_2^2$ becomes less
sensitive to the size of the core.  The main reason for this result is
that the overlap between the $Y_2^2$ potential and the inertial normal
modes of the full polytrope increases as $n$ increases, so that a
significant tidal response (albeit concentrated into narrow peaks) is
possible even in the absence of a core.

These results are in broad agreement with the findings of
\citet{2010MNRAS.407.1631P}, who calculated the energy transferred
(predominantly to inertial waves) to a slowly rotating $n=1$ polytrope
during a distant parabolic encounter.  They found that the energy
transferred via $Y_2^2$ was increased by nearly an order of magnitude
when a (perfectly rigid) solid core with $\alpha=0.5$ was introduced,
but hardly at all when $\alpha=0.25$.  This is broadly in line with
the red curve in Figure~\ref{f:int_l2_m2_various}; although a
parabolic encounter provides a tidal force that is localized in time,
their problem is not identical to the impulse problem considered in
this paper.  For example, the Fourier transform of the Heaviside
function has a singularity at zero frequency that is not present in
the case of a parabolic encounter.

\section{Conclusions}

In this paper we have examined the linear response of slowly rotating,
neutrally stratified astrophysical bodies to low-frequency tidal
forcing.  This problem may be regarded as a simplified model of tides
in convective regions of stars and giant planets.  The response can be
separated into non-wavelike and wavelike parts, where the former is
related instantaneously to the tidal potential and the latter may
involve resonances or other singularities.  The imaginary part of the
potential Love number of the body, which is directly related to the
rates of energy and angular momentum exchange in the tidal interaction
and to the rate of dissipation of energy, may have a complicated
dependence on the tidal frequency.  However, a certain
frequency-average of this quantity is independent of the dissipative
properties of the fluid and can be determined by means of an impulse
calculation.  The result is a strongly increasing function of the size
of the core when the tidal potential is a sectoral harmonic ($|m|=l$),
such as the $l=m=2$ potential usually considered, especially when the
body is not strongly centrally condensed.  However, the same is not
true for tesseral harmonics ($|m|<l$), which receive a richer response
and may therefore be important in determining tidal evolution even
though they are usually subdominant in the expansion of the tidal
potential.  We have also discussed analytically the low-frequency
response of a slowly rotating homogeneous fluid body to tidal
potentials proportional to spherical harmonics of degrees less than
five.  Tesseral harmonics of degrees greater than two, such as are
present in the case of a spin-orbit misalignment, can resonate with
inertial modes of the full sphere, leading to an enhanced tidal
interaction.  Similar behaviour can be expected in more realistic
models.

The calculations carried out in this paper are based on linear theory
and employ highly simplified interior structure models.  There are
various reasons why the complicated details of the frequency-dependent
response functions might not be applicable to astrophysical bodies.
The propagation of inertial waves may be impeded by nonlinearity
(leading to instability, wave breaking, etc.), or interrupted by
interactions with convection, magnetic fields, differential rotation,
buoyancy forces, etc.  The waves may not reflect perfectly from the
boundaries of the convective regions that support them.  These
complications are likely to tend to wash out some of the structure in
the frequency-dependent response curves.  Nevertheless, the integrated
response functions calculated in this paper are likely to be much more
robust; they can be related, as we have seen, to the energy
transferred in an impulsive interaction, which is independent of the
detailed way in which the waves subsequently propagate, reflect and
dissipate.  The normal-mode resonances that we have identified with
higher-order tidal potentials are also likely to be important even in
the presence of effects such as convection.

Tidal interactions at low orbital eccentricity can be studied by
analysing the tidal potential into a small number of components with a
harmonic dependence on time.  In this case the integrated responses
discussed in this paper can be taken as indicative of the typical
level of tidal dissipation, neglecting the complicated
frequency-dependence associated with the propagation of inertial waves
in a spherical shell.  Of particular importance are the scaling of the
imaginary part of the Love number (or the reciprocal of the tidal
quality factor) with the square of the dimensionless rotation rate
$\epsilon$, and its dependence on the size of a solid or fluid core
that excludes the inertial waves by total or partial reflection.  At
high orbital eccentricity, tidal interactions are more impulsive in
character because the tidal force is strongly peaked near pericentre,
and the integrated responses are more directly applicable to such
situations.

Previous works investigating tidally forced inertial waves in
astrophysical bodies have tended to emphasize either global normal
modes or the singular phenomena associated with wave attractors and
critical latitudes.  This paper attempts to bridge the gap between
these descriptions.  We consider that inviscid inertial waves in
spherical (or spheroidal) geometry form true normal modes only under
special conditions, i.e.\ when a core is absent and when the density
profile is sufficiently smooth.  We find classical resonances with
normal modes in coreless homogeneous bodies and polytropes.  As a core
is introduced and increased in size, some `memory' of these modes is
retained initially, but later the singular wave phenomena dominate.

The application of these findings to extrasolar planets and
solar-system bodies requires further work.  The interior structure of
giant planets is still quite uncertain; there may be non-smooth
features such as discontinuities in the density or its radial
derivative.  Progress in the modelling of planetary interiors will
certainly assist in the determination of their tidal responses.

\section*{acknowledgments}

This research was supported by STFC.  I am grateful to Jeremy Goodman
for helpful discussions at an early stage in this work.  I thank Pavel
Ivanov and the referee, Michel Rieutord, for their comments and
suggestions.

\appendix

\newpage

\onecolumn

\section{Responses of a homogeneous sphere}
\label{a:homogeneous}

In this section we give some explicit solutions for the linear
response of a slowly rotating homogeneous fluid body without a solid
core to low-frequency tidal forcing.  The notation follows that of
\citet{2009MNRAS.396..794O}, with a time-dependence $\rme^{-\rmi\omega
  t}$ in the frame rotating with the body, but viscous and frictional
damping forces are omitted, so $\omega_n=\omega+2m\Omega/[n(n+1)]$.

Given $\Psi=A(r/R)^lY_l^m$, the solution of the Helmholtz-like
equation is, as in Section~\ref{s:homogeneous}, $\Phi'=B(r/R)^lY_l^m$
for $r<R$ and $\Phi'=B(R/r)^{l+1}Y_l^m$ for $r>R$, with $2(l-1)B=3A$.
Then $X=C(r/R)^lY_l^m$ with $C=(A+B)R/(lg)$ and $g=GM/R^2$.  This
gives $\xi_{\rmnw,r}=aR(r/R)^{l-1}Y_l^m$, where $a=-lC/R^2$ is the
dimensionless non-wavelike radial tidal amplitude at the surface.  The
non-wavelike velocity is given by
\begin{equation}
  a_l=-\rmi\omega aR\left(\f{r}{R}\right)^{l-1},\qquad
  b_l=-\rmi\omega\f{a}{l}\left(\f{r}{R}\right)^{l-2}.
\end{equation}
The wavelike velocity solutions in various cases are as follows.

\subsection{The case $l=2$}

\begin{equation}
  c_1=\f{5aq_2\omega\Omega}{2\omega_1},\qquad
  W_0=\f{5aq_1q_2\omega\Omega^2}{\omega_1}r^2,\qquad
  W_2=\f{a\omega\Omega(m\omega_1-5q_2^2\Omega)}{2\omega_1}r^2.
\end{equation}
Note that a constant $c_1$, as occurs here, implies a special type of
solution, not really an inertial wave.  If $m=2$ this term vanishes
because then $q_2=0$.  If $m=1$ the term $c_1$ is a spin-over mode
(tilting the rotation axis).  If $m=0$ it is a spin-up mode (changing
the rotation rate).  The response is resonant when the denominator
vanishes, i.e.\ when $\omega+m\Omega=0$, which means that the tidal
frequency in the inertial frame vanishes.

\subsection{The case $l=3$}

\begin{equation}
  a_1=\f{14\rmi aq_2q_3\omega\Omega^2}{3(\omega_1\omega_2-3q_2^2\Omega^2)}\f{(R^2-r^2)}{R},\qquad
  b_1=\f{14\rmi aq_2q_3\omega \Omega^2}{3(\omega_1\omega_2-3q_2^2\Omega^2)}\f{(R^2-2r^2)}{Rr},\qquad
  c_2=\f{14aq_3\omega\omega_1\Omega}{9(\omega_1\omega_2-3q_2^2\Omega^2)}\f{r}{R}.
\end{equation}
\begin{equation}
  W_1=-\f{14aq_2q_3\omega(\omega+2m\Omega)\Omega^2}{3(\omega_1\omega_2-3q_2^2\Omega^2)}\f{r(R^2-r^2)}{R},\qquad
  W_3=a\omega\Omega\left[2m-\f{56q_3^2\omega_1\Omega}{3(\omega_1\omega_2-3q_2^2\Omega^2)}\right]\f{r^3}{9R}.
\end{equation}
The response is resonant when the common denominator vanishes, i.e.\ when
\begin{equation}
  \omega=\f{2}{3}\Omega\left[-m\pm\left(\f{9-m^2}{5}\right)^{1/2}\right],
\end{equation}
but not for $|m|=3$, because then $q_3$ in the numerator also vanishes.

\subsection{The case $l=4$}

\begin{equation}
  a_2=\f{81\rmi aq_3q_4\omega\omega_1\Omega^2}{9\omega_1\omega_2\omega_3-(27q_2^2\omega_3+32q_3^2\omega_1)\Omega^2}\f{r(R^2-r^2)}{R^2},\qquad
  b_2=\f{27\rmi aq_3q_4\omega\omega_1\Omega^2}{9\omega_1\omega_2\omega_3-(27q_2^2\omega_3+32q_3^2\omega_1)\Omega^2}\f{(3R^2-5r^2)}{2R^2},
\end{equation}
\begin{equation}
  c_1=-\f{81aq_2q_3q_4\omega\Omega^3}{9\omega_1\omega_2\omega_3-(27q_2^2\omega_3+32q_3^2\omega_1)\Omega^2}\f{(5R^2-7r^2)}{2R^2},\qquad
  c_3=\f{81aq_4\omega\Omega(\omega_1\omega_2-3q_2^2\Omega^2)}{9\omega_1\omega_2\omega_3-(27q_2^2\omega_3+32q_3^2\omega_1)\Omega^2}\f{r^2}{8R^2},
\end{equation}
\begin{equation}
  W_0=-\f{81aq_1q_2q_3q_4\omega\Omega^4}{9\omega_1\omega_2\omega_3-(27q_2^2\omega_3+32q_3^2\omega_1)\Omega^2}\f{r^2(10R^2-7r^2)}{2R^2},\qquad
  W_2=-\f{81aq_3q_4\omega\Omega^2(\omega_1^2-5q_2^2\Omega^2)}{9\omega_1\omega_2\omega_3-(27q_2^2\omega_3+32q_3^2\omega_1)\Omega^2}\f{r^2(R^2-r^2)}{2R^2},
\end{equation}
\begin{equation}
  W_4=a\omega\Omega\left[2m-\f{243q_4^2\Omega(\omega_1\omega_2-3q_2^2\Omega^2)}{9\omega_1\omega_2\omega_3-(27q_2^2\omega_3+32q_3^2\omega_1)\Omega^2}\right]\f{r^4}{16R^2}.
\end{equation}
Again, the response is resonant when the common denominator vanishes, i.e.\ when $\omega$ satisfies the cubic equation
\begin{equation}
  42\omega^3+63m\omega^2\Omega-36(2-m^2)\omega\Omega^2-4m(11-2m^2)\Omega^3=0,
\end{equation}
but not for $|m|=4$, because then $q_4$ in the numerator also vanishes.

\section{Analytical results for integrated responses}
\label{a:analytical}

\subsection{Homogeneous body with a solid core}

The equivalent of equation~(\ref{int_l2_m2_homogeneous}) for general
values of $l$ and $m$ is
\begin{equation}
  \int_{-\infty}^\infty\mathrm{Im}[K_l^m(\omega)]\,\f{\rmd\omega}{\omega}=\f{3(2l+1)\pi\epsilon^2[(l+1)^3(l-1)(2l+3)(l^2-m^2)+l^3(l+2)(2l-1)((l+1)^2-m^2)\alpha^{2l+1}]}{l^3(l+1)^3(l-1)^2(2l+3)(2l-1)(1-\alpha^{2l+1})}.
\end{equation}

\subsection{Homogeneous fluid body with a solid core of a different
  density}

The equivalent result in the case $l=m=2$ when the solid core has a
density $f^{-1}$ times that of the fluid envelope is
\begin{equation}
  \int_{-\infty}^\infty\mathrm{Im}[K_2^2(\omega)]\,\f{\rmd\omega}{\omega}=\f{100\pi}{63}\epsilon^2\left(\f{\alpha^5}{1-\alpha^5}\right)\left[1+\left(\f{1-f}{f}\right)\alpha^3\right]\left[1+\f{5}{2}\left(\f{1-f}{f}\right)\alpha^3\right]^{-2}.
\end{equation}
We recover equation~(\ref{int_l2_m2_homogeneous}) either as $f\to1$ or in the limit $\alpha^3\ll f$.

\subsection{Piecewise-homogeneous fluid body}

The equivalent result for the piecewise-homogeneous fluid model is
\begin{eqnarray}
  \int_{-\infty}^\infty\mathrm{Im}[K_2^2(\omega)]\,\f{\rmd\omega}{\omega}&=&\f{100\pi}{63}\epsilon^2\left(\f{\alpha^5}{1-\alpha^5}\right)(1-f)^2(1-\alpha)^4\left(1+2\alpha+3\alpha^2+\f{3}{2}\alpha^3\right)^2\left[1+\left(\f{1-f}{f}\right)\alpha^3\right]\nonumber\\
  &&\times\left[1+\f{3}{2}f+\f{5}{2f}\left(1+\f{1}{2}f-\f{3}{2}f^2\right)\alpha^3-\f{9}{4}(1-f)\alpha^5\right]^{-2}.
\label{pwh}
\end{eqnarray}
For small $f$ and small $\alpha$, specifically $\alpha^3\ll f\ll1$, we
again recover equation~(\ref{int_l2_m2_homogeneous}).

\label{lastpage}

\end{document}